\documentclass[sigconf]{acmart}
\usepackage{multirow} 
\usepackage{balance}
\usepackage{graphicx}
\usepackage{amsmath}
\usepackage{array}
\usepackage{amsfonts}
\usepackage{float}
\usepackage{color}
\usepackage{colortbl}
\usepackage{lipsum}
\AtBeginDocument{%
  \providecommand\BibTeX{{%
    \normalfont B\kern-0.5em{\scshape i\kern-0.25em b}\kern-0.8em\TeX}}}

\newcommand\blfootnote[1]{%
\begingroup 
\renewcommand\thefootnote{}\footnote{#1}%
\addtocounter{footnote}{-1}%
\endgroup 
}
\setcopyright{acmcopyright}
\copyrightyear{2022}
\acmYear{2022}
\acmConference[SIGIR '22] {Proceedings of the 45th International ACM SIGIR Conference on Research and Development in Information Retrieval}{July 11--15, 2022}{Madrid, Spain.}
\acmPrice{15.00}
\acmISBN{978-1-4503-8732-3/22/07}
\acmDOI{10.1145/3477495.3531715}

\begin{document}
\fancyhead{}
\title{Where Does the Performance Improvement Come From? \protect\\ - A Reproducibility Concern about Image-Text Retrieval}


\author{Jun Rao}
\authornote{Equal contributions from both authors. Work was done when Jun and Fei were interning at JD Explore Academy.}
\affiliation{%
  \institution{Harbin Institute of Technology, Shenzhen}
  \country{}
}

\author{Fei Wang}
\authornotemark[1]
\affiliation{%
  \institution{China University of Petroleum \protect\\ (East China)}
  \country{}
}

\author{Liang Ding}
\authornote{Corresponding Authors.}
\affiliation{%
  \institution{JD Explore Academy}
  \country{}
}
\email{dingliang1@jd.com}

\author{Shuhan Qi}
\authornotemark[2]
\affiliation{%
  \institution{Harbin Institute of Technology, Shenzhen;}
  \institution{Peng Cheng Laboratory}
  \country{}
}
\email{shuhanqi@cs.hitsz.edu.cn}

\author{Yibing Zhan}
\affiliation{%
  \institution{JD Explore Academy}
  \country{}
}

\author{Weifeng Liu}
\affiliation{%
  \institution{China University of Petroleum \protect\\ (East China)}
  \country{}
}

\author{Dacheng Tao}
\affiliation{%
  \institution{JD Explore Academy}
  \country{}
}



\begin{abstract}
This article aims to provide the information retrieval community with some reflections on recent advances in retrieval learning by analyzing the reproducibility of image-text retrieval models.
Due to the increase of multimodal data over the last decade, image-text retrieval has steadily become a major research direction in the field of information retrieval.
Numerous researchers train and evaluate image-text retrieval algorithms using benchmark datasets such as MS-COCO and Flickr30k.
Research in the past has mostly focused on performance, with multiple state-of-the-art methodologies being suggested in a variety of ways.
According to their assertions, these techniques provide improved modality interactions and hence more precise multimodal representations. In contrast to previous works, we focus on the reproducibility of the approaches and the examination of the elements that lead to improved performance by pretrained and nonpretrained models in retrieving images and text.

To be more specific, we first examine the related reproducibility concerns and explain why our focus is on image-text retrieval tasks. Second, we systematically summarize the current paradigm of image-text retrieval models and the stated contributions of those approaches. Third, we analyze various aspects of the reproduction of pretrained and nonpretrained retrieval models. To complete this, we conducted ablation experiments and obtained some influencing factors that affect retrieval recall more than the improvement claimed in the original paper. Finally, we present some reflections and challenges that the retrieval community should consider in the future. Our source code is publicly available at https://github.com/WangFei-2019/Image-text-Retrieval.


\end{abstract}



\maketitle
  
\section{Introduction}
As technology progresses, the content of information retrieval has evolved from a single-modality approach to a multimodal one~\cite{DBLP:conf/sigir/HuZPL19}. The continuous development of social platforms has resulted in an increase in the quantity of multimedia data on the internet, such as images and text. Finding similar content within such massive quantities of multimedia data has become a significant issue in the industry~\cite{DBLP:conf/sigir/GaoJCQLWHW20/fashionbert}. 
Due to the requirements of the practical applications, developing an effective image-text retrieval system has become a significant area of research of information retrieval.
The specific goal is to provide a flexible retrieval experience~\cite{DBLP:conf/cikm/RaoQQW0021} that indexes semantically relevant instances from one modality to another.

Image-text retrieval has been intensively investigated in recent years and can be divided into two categories according to whether using pretrained models.
On the one hand, \textit{visual-and-language pretraining} (VLP) based on pretrain-finetune paradigm has achieved state-of-the-art results on a range of downstream tasks such as image retrieval, visual question answering, and visual reasoning (\textit{e.g.}, \citet{vilbert,uniter}).
Most of these VLP models extend BERT~\cite{bert} to learn representations grounded in both visual and textual contexts. 
These VLP models mainly differ in designing the pretraining tasks, modality interaction, and the quantity of pretraining data~\cite{DBLP:conf/icml/JiaYXCPPLSLD21/scaling}.
Although these VLP models have been proposed and reported state-of-the-art results on various downstream tasks, there is still little research on what factors affect the final downstream task. 
To address this gap, we focus on the image-text retrieval task and attempt to compare these VLPs, to the best of our ability, exploring salient factors that may affect retrieval results. 
Additionally, while reproducing these VLP models, we raise concerns and think about the reproducibility of the results. On the other hand, current \textit{nonpretrained image-text retrieval models} are also a research hotspot because they generally require significantly fewer parameters compared to their pretrained counterparts, with the sacrifice of the performance~\cite{DBLP:conf/acl/LuZL20}.
Numerous methods \cite{SCAN,vse++,zhan2018comprehensive,yu2021deep,DBLP:conf/cvpr/caan,Wang_2019_ICCV/camp,DBLP:conf/aaai/DiaoZML21/SGRAF,DBLP:conf/iccv/LiZLLF19/VSRN} have been proposed in recent years, most of which claim to achieve better modality interactions and thus better multimodal representations. 
It is relatively easy to disentangle the factors that influence these nonpretrained models compared to pretrained models. 
We, therefore, chose a group of open-source methods, tried our best to reproduce the results of the original paper, and performed methods with different experimental setups to obtain new findings and the key factors that may influence the results.


We conducted experiments on two of the most widely used large-scale datasets, Flickr30k~\cite{DBLP:journals/tacl/YoungLHH14} and MS-COCO~\cite{DBLP:conf/eccv/LinMBHPRDZ14}. We tried 5 pretrained retrieval models and 6 nonpretrained retrieval models and reproduced these methods as closely as possible according to the papers' description and the provided codes. We conducted three separate experiments on both datasets, each with a different random seed, and took the final mean as the result reported in our table. Surprisingly, simple differences in initializations, hard samples, and seemingly insignificant details can result in dramatic differences in model performance. Moreover, we tried to run another ten experiments using nonpretrained methods with different random seeds on the Flickr30k dataset and draw the violin figure to show stability of these nonpretrained methods. 

In summary, our contributions in this paper are as follows:
\begin{itemize}
    \item We give a comprehensive overview of image-text retrieval learning methods, including modality embedding, modality interaction, similarity modeling, and a family of retrieval methods with pretrained and nonpretrained.
    \item We conduct a series of controlled studies in two benchmark datasets, raise concerns about the reproducibility of the settings of pretrained models, and discover that the improvements of nonpretrained models may come from hyperparameters, hard negative sampling strategies, and modality interaction types.
    \item We discuss the conjectures and give recommendations and insightful guidance in the information retrieval area.
\end{itemize}

\section{The need for reproducible image-text retrieval}
\subsection{Image-text Retrieval}
From 2018 to the present, many research papers related to cross-modal retrieval have been presented at major conferences, such as CVPR, ICCV, ECCV, MM, SIGIR, ICML, etc. Meanwhile, some easy-to-practice and effective methods \cite{DBLP:conf/cvpr/00010BT0GZ18/bottom-up-attention,SCAN,uniter,vilbert,kim2021vilt} have been widely used in practical commercial applications. 
With the growth of the Internet, the forms of multimodal data, such as photos, texts, audio, and videos, have expanded rapidly, with images and texts being the two most common modalities. As a result, how to retrieve these two fundamental modalities of vision and text is crucial and inspiring for more and different modalities retrieval. 

Image-text retrieval focuses on obtaining a set of sentences given a query image (image-to-text retrieval) and identifying images from candidates given a caption that describes their content (text-to-image retrieval). 
A major challenge of image-text retrieval is the need to model the semantic information of different modalities and align the semantic information of different modalities. 

Many current image-text retrieval methods encode the features of different modalities into a semantic space through modality-independent encoders and perform modal fusion to obtain the corresponding fusion features. Finally, the fusion features are converted into a similarity score to measure the similarity of the image and text by a head pooler. Following the completion of learning, the features of database items are calculated and indexed so that the retrieval system can efficiently perform retrieval similarity calculations to return the retrieval ranking results to the user.

\subsection{Reproducibility}
A remarkable series \cite{bert,DBLP:conf/nips/VaswaniSPUJGKP17,DBLP:conf/cvpr/HeZRS16/resnet,DBLP:conf/cvpr/00010BT0GZ18/bottom-up-attention,FASTERRCNN} of empirical successes in academia and industry \cite{DBLP:conf/sigir/GaoJCQLWHW20/fashionbert} has accompanied and nourished the rapid increase in academic research on image-text retrieval.
Through complicated module and model ensembles, extra parameter settings are provided to achieve performance benefits on datasets. 
These approaches are not very generic or useful, and it is difficult to maintain their effectiveness when circumstances change. However, proposals that are eventually embraced by the information retrieval community and practitioners are those that steadily increase performance across a wide range of "real-world" situations. An influential method should be highly generalizable and capable of many different parameter settings, such as transformers \cite{DBLP:conf/nips/VaswaniSPUJGKP17}, residual networks \cite{DBLP:conf/cvpr/HeZRS16/resnet}, and the newly proposed ViTAE~\cite{vitae,vitaev2}.  As a result, it is crucial to determine which approaches are reproducible and can be generalized in different settings and environments.

\section{A Unified Framework OF image-text retrieval}
\begin{figure*}[t]
    \centering
    \includegraphics[width=1.0\linewidth]{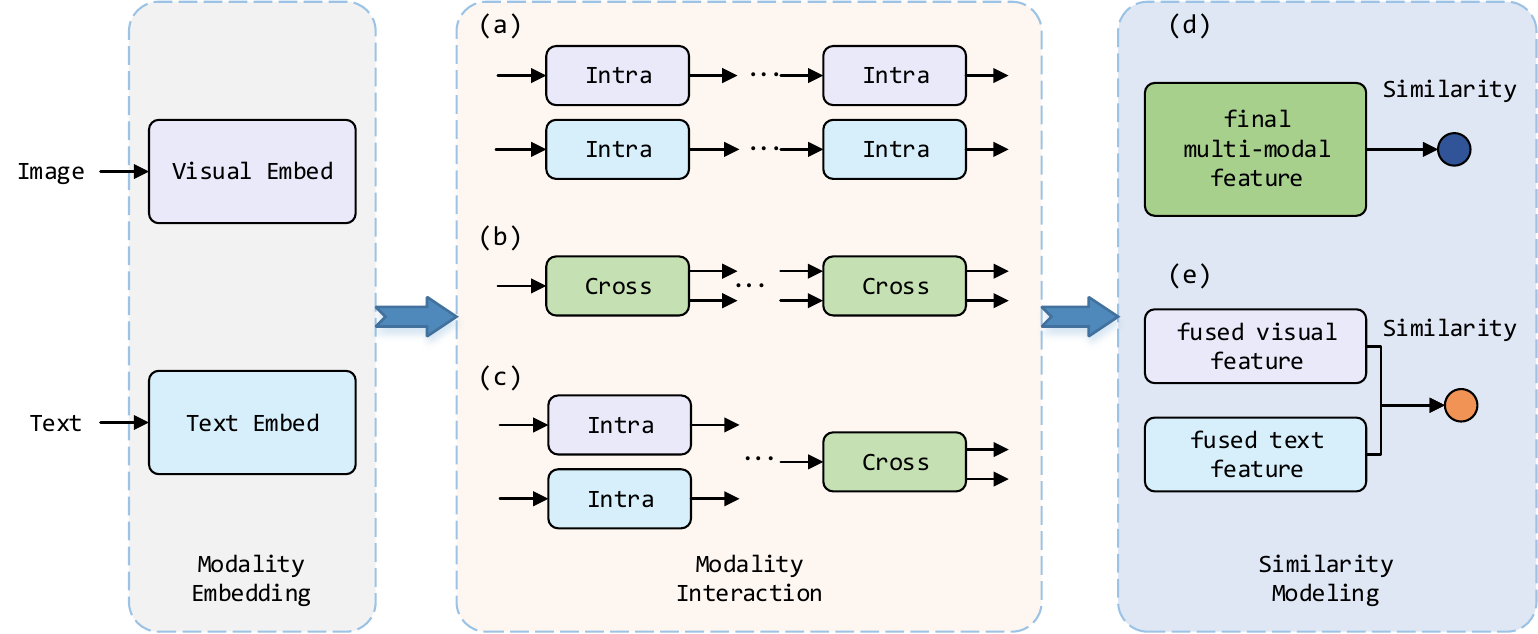}
    \caption{Overview image-text retrieval framework}
    \label{fig:overview}    
\end{figure*}
As shown in Figure  \ref{fig:overview}, we summarize the general process of the current image-text retrieval model and roughly divide each component of the retrieval model into three blocks, namely, modality embedding, modality interaction, and similarity calculation. In the following subsections, we describe the three key components and provide an architectural overview of image-text retrieval.

\subsection{Modality embedding}
The majority of work is devoted to enhancing the model's capability through modifying visual features, while text features are rarely considered~\cite{li2020unimo}. Most researchers have previously concentrated on visual features, thinking them to be the bottleneck affecting the retrieval model. However, we believe that learning how to use text features is also critical. Next, we demonstrate a series of visual and textual feature advancements.
\subsubsection{Visual representations}
\ \\
\textbf{Region feature.}
Region features are dominantly utilized among image-text retrieval models~\cite{vilbert,vl-bert,uniter,unicoder,oscar}. They are pretrained on the Visual Genome (VG) dataset processed by \cite{DBLP:conf/cvpr/00010BT0GZ18/bottom-up-attention} to obtain an off-the-shelf object detector, such as Faster R-CNN \cite{FASTERRCNN}.
The region feature extractor model can be varied by using different detection architectures, such as FPN \cite{DBLP:conf/cvpr/LinDGHHB17/fpn} and C4~\cite{DBLP:conf/cvpr/00010BT0GZ18/bottom-up-attention} or using different CNN backbones, such as ResNet101 \cite{vilbert,vl-bert} and ResNet152 \cite{DBLP:journals/corr/abs-1908-03557/visualbert,oscar}.
 Although features can greatly affect retrieval performance, previous work seems to be very tolerant of visual embedding. Even if the encoders are different, many methods still only compare the final retrieval accuracy and \textit{\textbf{do not mention the effect}} of visual embedding on their own model. Moreover, the number of regions also has a great impact on the final result. However, some methods \cite{oscar,vinvl} use more regions, resulting in unfair comparisons.
\\
\textbf{Grid feature and patch projection.}
These two types of features are mostly used in the pretrained image-text retrieval model and are rarely used in the nonpretrained model because of their worse retrieval performance. Nevertheless, once pretrained with a large amount of image-text pairs, these two types of features seem to be effective and meaningful. The grid feature was first proposed in the VQA task by \citet{DBLP:conf/cvpr/JiangMRLC20} to reduce the slow region selection operation. The grid feature is also extracted through the pretrained CNN model. Compared with region features, grid features do not need region selection, and thus using grid features is faster in practice. Patch Projection \cite{dosovitskiy2020image} was first adopted in image-text retrieval by ViLT \cite{kim2021vilt}. Compared with the previous two types of features, using patch projection feature is more direct and faster with less parameter consumption, without region selection or pretrained CNN. However, in practice, the performance of recall when using the grid feature or the patch projection is still worse than when using regional features for image-text retrieval.

\subsubsection{Textual representations}
\ \\
Different from visual representation, text representation does not seem to have great differences. Most methods directly use the powerful pretrained language model Bert \cite{bert} or GRU \cite{DBLP:journals/tsp/SchusterP97/bigru,DBLP:journals/corr/BahdanauCB14/gru} to obtain sentence dense embedding,   while ignoring the multi-granularity textual representations of sequential information, phrase information, lexical information, and noun information~\cite{bm25,ding2021progressive}. However, these items all play important roles for text retrieval~\cite{DBLP:conf/mm/Qu0CN020/campera,DBLP:conf/ijcai/HuLLYC19}. Moreover, current image-text retrieval lacks a discussion of the use of such textual information to get strong textual representations. Borrowing the success from the multi-lingual~\cite{conneau2019cross,ding2020self,wu2021slua,zan2022bridging} may be a potential direction.

\subsection{Modality interaction}
Most nonpretrained models ~\cite{DBLP:conf/iccv/LiZLLF19/VSRN,DBLP:conf/aaai/DiaoZML21/SGRAF,DBLP:conf/mm/Qu0CN020/campera,SCAN} claim that their contribution includes better modality interactions. Modality interactions can be roughly divided into two basic categories, as shown in Figure  \ref{fig:overview}. Mode (a) is self-interaction, which usually uses the attention mechanism to interact with the features in the model or just uses the embedding of the modality encoder. The second mode, as shown in (b), is the interaction between modalities. Usually, different modal features aggregate and share features through different attention mechanisms, such as graph attention networks \cite{DBLP:conf/iclr/VelickovicCCRLB18/GAT}, self-attention \cite{DBLP:journals/corr/BahdanauCB14/gru}, and co-attention \cite{vilbert}. The third mode (c) is the combination of the first two. Better retrieval results can be obtained through artificially defined feature interactions.

\subsection{Similarity modeling}
Similarity modeling can be roughly divided into two categories, as shown in Figure  \ref{fig:overview} (d) and (e).
The first category (d) obtains the joint representation of the image-text pair after multimodal interaction and usually appends a fully connected (FC) layer to obtain the similarity followed by softmax to predict a two-class probability $p^{itm}$. Many VLP models adopt the image-text matching (ITM) loss, which predicts whether an image and text pair match:
\begin{equation}\label{itm}
\mathcal{L}_{\mathrm{itm}}=\mathbb{E}_{(I, T) \sim D} \mathrm{H}\left({y}^{\mathrm{itm}}, {p}^{\mathrm{itm}}(I, T)\right),
\end{equation}
where $y^{itm}$ is a 2-dimensional one-hot vector representing the ground-truth label, and $\mathrm{H}$ is the cross-entropy. In general, the calculation of Equation (\ref{itm}) is usually used in the pretrained retrieval model.

The second method (e) obtains the representation of each modality and calculates the similarity between unimodal features by directly exploiting or learning a similarity function. This type of similarity modeling method usually adopts contrastive image-text matching losses, which have been successful in self-supervised representation learning \cite{DBLP:journals/corr/abs-1807-03748/Contrastive}.
For each image and text, the processes for calculating the softmax-normalized image-to-text and text-to-image similarity is defined as follows:
\begin{equation}\label{i2t}
p_{m}^{\mathrm{i} 2 \mathrm{t}}(I)=\frac{\exp \left(s\left(I, T_{m}\right) / \tau\right)}{\sum_{m=1}^{M} \exp \left(s\left(I, T_{m}\right) / \tau\right)},
\end{equation}
\begin{equation}\label{t2i}
p_{m}^{\mathrm{t} 2 \mathrm{i}}(T)=\frac{\exp \left(s\left(T, I_{m}\right) / \tau\right)}{\sum_{m=1}^{M} \exp \left(s\left(T, I_{m}\right) / \tau\right)},
\end{equation}
where $\tau$ is the temperature parameter. Let ${y}^{\mathrm{i} 2 \mathrm{t}}(I)$ and
${y}^{\mathrm{t} 2 \mathrm{i}}(T)$ 
denote the ground truth, where the score equals to 1 if matched and otherwise 0. Then, the contrastive loss can be defined as follows:
\begin{equation}\label{itcl}
\mathcal{L}_{\mathrm{itc}}=\frac{1}{2} \mathbb{E}_{(I, T) \sim D}\left[\mathrm{H}\left({y}^{\mathrm{i} 2 \mathrm{t}}(I), {p}^{\mathrm{i} 2 \mathrm{t}}(I)\right)+\mathrm{H}\left({y}^{\mathrm{t} 2 \mathrm{i}}(T), {p}^{\mathrm{t} 2 \mathrm{i}}(T)\right)\right].
\end{equation}

Another simple form for updating similarity, comparable to the contrastive loss, is the bidirectional ranking loss, as illustrated in Equation  (\ref{hard_hinge_loss}):

\begin{equation}\label{hard_hinge_loss}
\begin{array}{r}
\mathcal{L}(\boldsymbol{I}, \boldsymbol{T})=\sum\left[\mu-{s}(\boldsymbol{I}, \boldsymbol{T})+{s}\left(\boldsymbol{I}, \boldsymbol{T}^{-}\right)\right]_{+} \\
+\sum\left[\mu-{s}(\boldsymbol{I}, \boldsymbol{T})+{s}\left(\boldsymbol{I}^{-}, \boldsymbol{T}\right)\right]_{+}.
\end{array}
\end{equation}

Compared to a pairwise loss, VSE++ \cite{vse++} employs batch hard-negative mining to increase embedding flexibility and make optimization easier:
\begin{equation}\label{hardest_hinge_loss}
\begin{array}{r}
\mathcal{L}(\boldsymbol{I}, \boldsymbol{T})=\max\left[\mu-{s}(\boldsymbol{I}, \boldsymbol{T})+{s}\left(\boldsymbol{I}, \boldsymbol{T}^{-}\right)\right]_{+} \\
+\max\left[\mu-{s}(\boldsymbol{I}, \boldsymbol{T})+{s}\left(\boldsymbol{I}^{-}, \boldsymbol{T}\right)\right]_{+},
\end{array}
\end{equation}
where $[x]+ = max(x, 0)$ is a clip function, $s(,)$ indicates the similarity prediction function, and $\mu$ is a positive constant, which we term the margin. In Equation (\ref{hardest_hinge_loss}), compared to a pairwise loss (Equation (\ref{hard_hinge_loss}) ), this loss addresses about the rank of the points with respect to a query rather than their exact distance, while another considers the sum of the violations for each negative sample.

The concept behind these functions is to increase the relevance score between an image and its corresponding text while decreasing the relevance score between an image and its irrelevant words.
In terms of repeatability, how training losses and samples are chosen can significantly impact the ultimate retrieval outcome.

\section{Materials and Methods}
\subsection{Datasets}
MS-COCO \cite{DBLP:conf/eccv/LinMBHPRDZ14} and Flickr30k \cite{DBLP:journals/tacl/YoungLHH14} have been used as benchmark datasets in most methods. The MS-COCO and Flickr30k datasets contain 123,287 and 31,783 images, respectively, and each image has five corresponding sentence descriptions. Most of the methods claim to split by \citet{DBLP:journals/pami/KarpathyF17}, using 121,287/1,000/1,000 images for training/validation/testing in Flickr30k
and 113,287/5,000/\\
5,000 images for training/validation/testing in MS-COCO dataset.
During the replication phase, however, we discovered that almost all algorithms combine the data from the validation set with the training data in order to generate higher test set results. Furthermore, because MS-COCO's 5k test is extremely time-consuming, some approaches employ the 1k test, which averages 5-fold of 1k test images from 5k images. However, it is still uncertain how to split and whether the result according to the split represents the best dividing outcome. As a result, for a more consistent comparison later, we use a unified division approach and average several measurements.

\subsection{Evaluation metrics}

At the test time, the result performance for image-text retrieval is reported by recall at K (R@K) which represents the  ranking proportion of ground-truth queries within the top K.  R@1, R@5, and R@10 are our evaluation metrics. To conveniently describe the experiment,  we abbreviate ``Image-to-text Retrieval'' and ``Text-to image Retrieval'' as ``IR'' and ``TR'', respectively.

\subsection{Models}\label{model_des}
\subsubsection{pretrained models}
\ \\
The pretrained language model has gained considerable interest from the natural language processing, computer vision, and information retrieval communities because it can use self-supervised learning through unified pre-training and performs well on many downstream tasks.
The visual-and-language pretraining (VLP) models achieve better performance in different downstream tasks. Most of the VLP models are pretrained on the image-text pairs of Google Conceptual Captions (GCC) \cite{DBLP:conf/acl/SoricutDSG18/gcc}, SBU Captions (SBU) \cite{DBLP:conf/nips/OrdonezKB11/sbu}, Microsoft COCO (MS-COCO) \cite{DBLP:conf/eccv/LinMBHPRDZ14} and VG datasets. Existing VLPs are frequently directed at a variety of downstream tasks, resulting in many VLPs that have not been evaluated on the image-text retrieval task. Therefore, it is meaningless to make comparisons with these methods without image-text retrieval results. We selected these VLPs based on the availability of full image-text retrieval test results and the influence of their citation count. These are the five models we chose:
ViLBERT \cite{vilbert},  PixelBERT \cite{DBLP:journals/corr/abs-2004-00849/pixel-bert}, Unicoder-VL  \cite{unicoder},  
UNITER \cite{uniter}, and ViLT \cite{kim2021vilt}.
We summarize these VLPs in Table \ref{vlp_compare}. These models share similar text encoders (BERT) and similar visual encoders (ROIs), but use different pre-training tasks and modality interaction architectures. On the downstream image-text retrieval tasks, these models all use the ITM loss to optimize multi-modal features of type (d), as shown in Equation  (\ref{itm}).

\textbf{ViLBERT \cite{vilbert}} introduced a co-attention mechanism to fuse the features of the visual and the text flow and obtained fused visual features and text features, respectively. 
This modality interaction method belongs to (c) in Figure  \ref{fig:overview}, which is also the most important contribution of this paper. This work is the originator of the pretrained visual-and-language model, and it has approximately 1,000 citations.
\textbf{PixelBERT \cite{DBLP:journals/corr/abs-2004-00849/pixel-bert}} feeds the text and image with CNN embeddings into the transformer together, which indicates the single-stream framework and belongs to type (b) in Figure  \ref{fig:overview}. It uses a multimodal transformer to align visual-and-language information and became the standard fusion method for the subsequent single-stream model. In addition,
it reports the complete image-text retrieval results but lacks the code and details of the implementation. \textbf{Unicoder-VL  \cite{unicoder}} uses a larger pre-training dataset (CC3M) and the contrast loss with the hardest in-batch negatives (Equation  (\ref{hardest_hinge_loss}) ) to optimize the image-text retrieval task for the first time. Nevertheless, neither the code nor the checkpoints for this project are open source. \textbf{PixelBERT} and \textbf{Unicoder-VL} lack code and details, it is basically impossible to reproduce the pretraining and downstream task results.
However, due to the influence and inspiration of these two works, we still consider these two methods in the subsequent discussion. \textbf{UNITER \cite{uniter}} and Unicoder-VL are basically the same architectures, belonging to the type (b) in Figure  \ref{fig:overview}. The difference is that UNITER uses a better combination of pretraining tasks and larger pretraining datasets. It also thoroughly examines the results on image-text retrieval datasets. Although this work provides training checkpoints and open-source code, it is nearly impossible to reproduce due to the hard negative mining time limit. This work models similarity (type d) using ITM loss (Equation  (\ref{itm}) ) as in previous work but with hard negative mining, which may improve nearly 5 $\sim$ 10 points in R@1. \textbf{ViLT\cite{kim2021vilt}} is one of the simplest VLP models. It is similar to UNITER in that it uses the same architectural type (b) and pretraining tasks, as well as uniformly input image patch and text encoding into the transformer to achieve competitive performance in the image-text retrieval task. Similar to UNITER, its similarity modeling (d) also uses ITM loss (Eq. (\ref{itm})). This method is easier to reproduce due to the simplicity and less extra setup.



\subsubsection{Nonpretrained models}
\ \\

Direct comparison of nonpretrained models and VLP is not fair due to the use of more data and longer training time. 
In the case of limited resources, it is also necessary to study nonpretrained models. The claimed main improvement of the nonpretrained models is mainly the modality interaction and similarity modeling in Figure  \ref{fig:overview}. 
Therefore, we use 6 nonpretrained models with open source code for experimental comparison to determine the extent to which these assumptions hold.
We show the differences in the architecture of these nonpretrained models in Table \ref{tab:non_vlp_setup}. The claimed contributions of the individual models are further explained next.

\textbf{VSE++ \cite{vse++}} includes the in-batch hard-negative mining technique in the ranking loss, which contributed significantly to the improvement as they claim. Additionally, unlike many later works, their visual encoding uses CNN, and text encoding uses GRU. VSE++ obtains the modal encoding of type (a), maps visual and text features to a representation space, and obtains the similarity of the two modalities through the dot product of (e).
\textbf{SCAN \cite{SCAN}} employs a stacked cross-attention model to predict similarity by taking into account the dense paired cross-modal interaction. Different from VSE++ \cite{vse++}, SCAN uses regions of interest (ROIs), to obtain the visual embedding. Then, SCAN uses the attention between modalities to obtain the fused modal information through the type (b) and obtains the final global image-text matching score by the mean of (d), as shown in Figure  \ref{fig:overview}.
\textbf{VSRN \cite{DBLP:conf/iccv/LiZLLF19/VSRN}} provides an interpretable and straightforward reasoning model by generating visual representations that capture significant items and semantic concepts in a picture. This technique focuses on interactions within visual modalities of (a). This demonstrates that the modal information of vision has not been fully exploited. Moreover, it applies the inner product as the similarity function in the joint embedding space, belonging to type (e).
\textbf{SAEM \cite{DBLP:conf/mm/WuWSH19/saem}} employs self-attention embeddings to take advantage of fragment relations in pictures or texts and aggregate fragment information into visual and textual embeddings. The modality interaction can be classified into (a). Similar to the four previous works, the basic loss used in the similarity modeling is a contrastive loss (Equation  (\ref{hardest_hinge_loss}) ). Furthermore, SAEM \cite{DBLP:conf/mm/WuWSH19/saem} adds hard negative mining on the angular loss \cite{DBLP:conf/iccv/WangZWLL17/Angular_loss} to model similarity of type (e).
\textbf{CAMERA \cite{DBLP:conf/mm/Qu0CN020/campera}} does not use a pair of image-text data for training but adds image-text joint training for multiview descriptions, and selects content information through an attention module, which takes advantage of intra-modal interactions (a). Although CAMERA also uses a contrastive loss similar to previous works to map features of different modalities into a representation space of type (e), CAMERA introduces a diversity regularization term that causes a difference in the loss term. This causes additional parameter adjustments and increases the difficulty for subsequent improvement exploration.
\textbf{SGRAF \cite{DBLP:conf/aaai/DiaoZML21/SGRAF}} designs the SGR module for graph reasoning and the SAF to filter useless information, using type (c) to conduct modality interaction and obtain better semantic alignment. It also uses a contrastive loss with the hardest negative (Equation  (\ref{hardest_hinge_loss}) ), using the method (d) to model similarity.

Although the authors of the corresponding studies assert that these models function well, there are still some problems and opportunities for improvement.
To begin, unlike VLPs, modality embeddings are investigated infrequently in nonpretrained models. Most approaches encode the image input using 36 visual regions and the text encoder GRU.
Second, these approaches are not generalizable and exhibit a high parameter sensitivity. SCAN \cite{SCAN}, VSRN \cite{DBLP:conf/iccv/LiZLLF19/VSRN}, and CAMERA \cite{DBLP:conf/mm/Qu0CN020/campera} report ensemble results. By doing so, these strategies improve reporting results but limit the method's generalizability and ease of use. Additionally, they rely excessively on the granularity of feature encoding and filter and weight modal features with varying granularities using customized fine-grained interaction modules.
Finally, the original publication poorly stated several critical parts of the models, although these elements are frequently critical for influencing the model's outcomes.


\section{Analysis}
We make tables of the experimental setup of all methods, as shown in Table \ref{vlp_compare} and Table \ref{tab:non_vlp_setup}.
\begin{table*}[h]
\small
\centering
\caption{Comparisons with existing VLP methods and details on image-text retrieval. {\textdagger} represents methods that cannot reproduce results due to lack of code and training details. }\label{vlp_compare}
\resizebox{\textwidth}{!}
{
\begin{tabular}{@{}l|l|l|l|l|l|ll|ll|l|l|l|l|l@{}}
\toprule
\multirow{2}{*}{Method}                                                     & \multirow{2}{*}{Params} & \multirow{2}{*}{Architecture}                                                                                                                                            & \multirow{2}{*}{\begin{tabular}[c]{@{}l@{}}Visual \\      Tokens\end{tabular}} & \multirow{2}{*}{\begin{tabular}[c]{@{}l@{}}Pre-train\\      Datasets\end{tabular}} & \multirow{2}{*}{\begin{tabular}[c]{@{}l@{}}Pre-train\\      Tasks\end{tabular}}          & \multicolumn{2}{l|}{Flickr30k}                                                                         & \multicolumn{2}{l|}{COCO}                                                                              & \multirow{2}{*}{BS} & \multirow{2}{*}{warmup \%} & \multirow{2}{*}{Loss}                                           & \multirow{2}{*}{tricks}                                                    & \multirow{2}{*}{code} \\ \cmidrule(lr){7-10}
                                                                            &                         &                                                                                                                                                                          &                                                                                &                                                                                    &                                                                                          & \multicolumn{1}{l|}{epoch}         & LR                                                                & \multicolumn{1}{l|}{epoch}         & LR                                                                &                     &                            &                                                                 &                                                                            &                       \\ \midrule
\begin{tabular}[c]{@{}l@{}}ViLBERT\\      (paper \cite{vilbert}/reproduction \cite{DBLP:conf/emnlp/ZhangHJIS20})\end{tabular} & 221M                    & \begin{tabular}[c]{@{}l@{}}one single-modal   Transformer\\      (language)\\      + one cross-modal Transformer\\      (with restricted attention pattern)\end{tabular} & image RoI                                                                      & CC                                                                                 & \begin{tabular}[c]{@{}l@{}}1) MLM\\      2) ITM\\      3) MIM\end{tabular}               & \multicolumn{1}{l|}{20/17}         & 4e-5                                                              & \multicolumn{1}{l|}{-/17}          & 4e-5                                                              & 64                  & 0.1                        & \begin{tabular}[c]{@{}l@{}}cross\\      entropy\end{tabular}    & \begin{tabular}[c]{@{}l@{}}1) HNM\\      2) FP16\end{tabular}              & PyTorch               \\ \midrule
PixelBERT{\textdagger} \cite{DBLP:journals/corr/abs-2004-00849/pixel-bert}                                                                  & 142M                    & single cross-modal Transformer                                                                                                                                           & CNN                                                                            & \begin{tabular}[c]{@{}l@{}}MS-COCO\\      VG\end{tabular}                           & \begin{tabular}[c]{@{}l@{}}1) MLM\\      2) ITM\end{tabular}                             & \multicolumn{1}{l|}{10}            & 1e-4                                                              & \multicolumn{1}{l|}{4}             & 1e-4                                                              & 512                 & -                          & \begin{tabular}[c]{@{}l@{}}cross\\      entropy\end{tabular}    & \begin{tabular}[c]{@{}l@{}}1) HNM\\      2) DA\\      3) FP16\end{tabular} & no                    \\ \midrule
Unicoder-VL{\textdagger} \cite{unicoder}                                                                & 110M                    & single cross-modal Transformer                                                                                                                                           & image RoI                                                                      & CC                                                                                 & \begin{tabular}[c]{@{}l@{}}1) MLM\\      2) ITM\\      3) MIM\end{tabular}               & \multicolumn{1}{l|}{-}             & 5e-5                                                              & \multicolumn{1}{l|}{-}             & 5e-5                                                              & 192                 & 0.1                        & \begin{tabular}[c]{@{}l@{}}contrastive\\      loss\end{tabular} & \begin{tabular}[c]{@{}l@{}}1) HNM\\      2) FP16\end{tabular}              & no                    \\ \midrule
\begin{tabular}[c]{@{}l@{}}UNITER\\      (paper \cite{uniter}/reproduction \cite{DBLP:conf/emnlp/ZhangHJIS20})\end{tabular}  & 110M                    & single cross-modal Transformer                                                                                                                                           & image RoI                                                                      & \begin{tabular}[c]{@{}l@{}}CC\\      SBU\\      MS-COCO\\      VG\end{tabular}      & \begin{tabular}[c]{@{}l@{}}1) MLM\\      2) ITM\\      3) MIM\\      4) WRA\end{tabular} & \multicolumn{1}{l|}{5000 steps/15} & \begin{tabular}[c]{@{}l@{}}5e-5\\      /\\      4e-5\end{tabular} & \multicolumn{1}{l|}{5000 steps/15} & \begin{tabular}[c]{@{}l@{}}5e-5\\      /\\      4e-5\end{tabular} & 8/64                & 0.1                        & \begin{tabular}[c]{@{}l@{}}cross\\      entropy\end{tabular}    & \begin{tabular}[c]{@{}l@{}}1) HNM\\      2) FP16\end{tabular}              & PyTorch               \\ \midrule
ViLT \cite{kim2021vilt}                                                                       & 111M                    & single cross-modal Transformer                                                                                                                                           & image patch                                                                    & \begin{tabular}[c]{@{}l@{}}CC\\      SBU\\      MS-COCO\\      VG\end{tabular}      & \begin{tabular}[c]{@{}l@{}}1) MLM\\      2) ITM\end{tabular}                             & \multicolumn{1}{l|}{15}            & 1e-4                                                              & \multicolumn{1}{l|}{10}            & 1e-4                                                              & 256                 & 0.1                        & \begin{tabular}[c]{@{}l@{}}cross\\      entropy\end{tabular}    & \begin{tabular}[c]{@{}l@{}}1) DA\\      2) FP16\end{tabular}               & PyTorch               \\ \bottomrule
\end{tabular}
}
\end{table*}
\begin{table*}[h]
\small
\centering
\caption{Comparisons with existing nonpretrained methods and details in image-text retrieval. The data corresponding to the column where LR is located is the initial learning rate$/$the epoch when the learning rate changes$/$the Change rate. The ``\textit{each}'' means that the change will occur after the specified number of epochs. }\label{tab:non_vlp_setup}
\resizebox{\textwidth}{!}{
\begin{tabular}{@{}cc|ccc|ccc|c|c|c|c|c|c@{}}
\toprule
\multicolumn{2}{c|}{\multirow{2}{*}{Method}}            & \multicolumn{3}{c|}{Flickr30k}                                        & \multicolumn{3}{c|}{MS-COCO}                                           & \multirow{2}{*}{Visual Encoder} & \multirow{2}{*}{Text Encoder} & \multirow{2}{*}{Framework} & \multirow{2}{*}{Loss}                                                                             & \multirow{2}{*}{Params} & \multirow{2}{*}{Cites} \\ \cmidrule(lr){3-8}
\multicolumn{2}{c|}{}                                   & \multicolumn{1}{c|}{Epoch} & \multicolumn{1}{c|}{BS} & LR     & \multicolumn{1}{c|}{Epoch} & \multicolumn{1}{c|}{BS} & LR     &                                 &                               &                            &                                                                                                   &                         &                        \\ \midrule
\multicolumn{2}{c|}{VSE++ \cite{vse++}}                        & \multicolumn{1}{c|}{30}    & \multicolumn{1}{c|}{128}        & 0.0002/15/$\times$0.1 & \multicolumn{1}{c|}{30}    & \multicolumn{1}{c|}{128}        & 0.0002/15/$\times$0.1 & CNN                             & GRU                           & a, e                          & contrastive loss                                                                                  & 67M                     & 610                    \\ \midrule
\multicolumn{2}{c|}{SCAN \cite{SCAN}}                         & \multicolumn{1}{c|}{30}    & \multicolumn{1}{c|}{128}        & 0.0002/15/$\times$0.1 & \multicolumn{1}{c|}{20}    & \multicolumn{1}{c|}{128}        & 0.0005/10/$\times$0.1 & image RoI                       & Bi-GRU                        & b, d                          & contrastive loss                                                                                  & 9M                      & 475                    \\ \midrule
\multicolumn{2}{c|}{VSRN \cite{DBLP:conf/iccv/LiZLLF19/VSRN}}                         & \multicolumn{1}{c|}{30}    & \multicolumn{1}{c|}{128}        & 0.0002/15/$\times$0.1 & \multicolumn{1}{c|}{30}    & \multicolumn{1}{c|}{128}        & 0.0002/15/$\times$0.1 & image RoI                       & Bi-LSTM                       & a, e                          & \begin{tabular}[c]{@{}c@{}}a hinge-based triplet\\ ranking loss, log-likelihood loss\end{tabular} & 140M                    & 162                    \\ \midrule
\multicolumn{1}{c|}{\multirow{2}{*}{SGRAF \cite{DBLP:conf/aaai/DiaoZML21/SGRAF}}} & SGR & \multicolumn{1}{c|}{40}    & \multicolumn{1}{c|}{128}        & 0.0002/30/$\times$0.1 & \multicolumn{1}{c|}{20}    & \multicolumn{1}{c|}{128}        & 0.0002/10/$\times$0.1 & \multirow{2}{*}{image RoI}      & \multirow{2}{*}{Bi-GRU}       & \multirow{2}{*}{c, d}         & \multirow{2}{*}{contrastive loss}                                                                 & 19M                     & \multirow{2}{*}{16}    \\ \cmidrule(lr){2-8} \cmidrule(lr){13-13}
\multicolumn{1}{c|}{}                             & SAF & \multicolumn{1}{c|}{30}    & \multicolumn{1}{c|}{128}        & 0.0002/20/$\times$0.1 & \multicolumn{1}{c|}{20}    & \multicolumn{1}{c|}{128}        & 0.0002/10/$\times$0.1 &                                 &                               &                            &                                                                                                   & 18M                     &                        \\ \midrule
\multicolumn{2}{c|}{SAEM \cite{DBLP:conf/mm/WuWSH19/saem}}                         & \multicolumn{1}{c|}{30}    & \multicolumn{1}{c|}{64}         & 0.0001/\textit{each}10/$\times$0.1 & \multicolumn{1}{c|}{30}    & \multicolumn{1}{c|}{64}         & 0.0001/\textit{each}10/$\times$0.1 & image RoI                       & BERT                          & a, e                          & contrastive loss and angular loss                                                                 & 114M                    & 40                     \\ 
\midrule
\multicolumn{2}{c|}{CAMERA \cite{DBLP:conf/mm/Qu0CN020/campera}}                       & \multicolumn{1}{c|}{30}    & \multicolumn{1}{c|}{128}        & 0.0001/\textit{each}10/$\times$0.1 & \multicolumn{1}{c|}{40}    & \multicolumn{1}{c|}{128}        & 0.0001/\textit{each}20/$\times$0.1 & image RoI                       & BERT                          & a, e                          & contrastive loss  and diversity regularization                                                    & 156M                    & 15                     \\
\bottomrule
\end{tabular}}
\end{table*}
\begin{table*}[h]
\small
\centering
\caption{Comparisons with existing VLP methods and their results in image-text retrieval. ``-'' represents the results of the original paper that were not given. {\textdagger} represents methods that cannot reproduce results due to the lack of code and details.}\label{vlp_result}
\resizebox{\textwidth}{!}
{
\begin{tabular}{@{}ccccccccccccc@{}}
\toprule
\multirow{2}{*}{Method}                                                         & \multicolumn{6}{c}{Flickr30k}                                       & \multicolumn{6}{c}{MS-COCO (5K)}                                           \\ \cmidrule(l){2-13} 
                                                                                & IR@1       & IR@5       & IR@10      & TR@1      & TR@5      & TR@10     & IR@1       & IR@5       & IR@10      & TR@1      & TR@5      & TR@10      \\ \midrule
\begin{tabular}[c]{@{}c@{}}ViLBERT\\      (paper/reproduction)\end{tabular}     & 58.2/59.1  & 84.9/85.7  & 91.5/92.0  & -/76.8    & -/93.7    & -/97.6    & -/38.6     & -/68.2     & -/79.0     & -/53.5    & -/79.7    & -/87.9     \\
\begin{tabular}[c]{@{}c@{}}PixelBERT\textdagger \\      (R50/X152)\end{tabular}             & 59.8/71.5  & 85.5/92.1  & 91.6/95.8  & 75.7/87   & 94.7/98.9 & 97.1/99.5 & 41.1/50.1  & 69.7/77.6  & 80.5/86.2  & 53.4/63.6 & 80.4/87.5 & 88.5/93.6  \\
Unicoder-VL\textdagger                                                                     & 71.5       & 90.9       & 94.9       & 86.2      & 96.3      & 99.0        & 46.7       & 76.0         & 85.3       & 62.3      & 87.1      & 92.8       \\
\begin{tabular}[c]{@{}c@{}}UNITER-Base\\      (paper/reproduction)\end{tabular} & 72.52/62.9 & 92.36/87.2 & 96.08/92.7 & 85.9/78.3 & 97.1/93.3 & 98.8/96.5 & 50.33/37.8 & 78.52/67.3 & 87.16/78.0 & 64.4/52.8 & 87.4/79.7 & 93.08/87.8 \\
\begin{tabular}[c]{@{}c@{}}ViLT-DA \\      (paper/reproduction)\end{tabular}     & 62.2/62.3  & 87.6/87.6  & 93.2/93.5  & 83.7/82.9 & 97.2/98.1 & 98.1/98.1 & 42.6/42.2  & 72.8/73.2  & 83.4/84.0  & 62.9/62.7 & 87.1/87.5 & 92.7/93.0  \\ \bottomrule
\end{tabular}
}
\end{table*}
\subsection{Pretrained Models}
We compare existing pretrained image-text retrieval models and present their detailed settings and parameter comparisons in Table \ref{vlp_compare}.
We analyze the ability of the retrieval model, the impact of the factor, and the reproducibility from the following two perspectives: the quantity of pretraining data and additional settings.
Although the VLP model is hard to make a fair comparison due to the differences aforementioned in section \ref{model_des}, we attempt to obtain some insightful conclusions considering reproducibility and practical improvement by comparing several models with the most similar settings.
\subsubsection{Concerning of pre-training}
\label{sec:data_size}\ \\
\textbf{\textit{As noted in \citet{DBLP:conf/icml/JiaYXCPPLSLD21/scaling}, it holds true that downstream tasks such as image-text retrieval 
perform better with more pretraining data.}} 
From the perspective of pretrained data, the pretrained data of the 5 models are divided into 3 categories. As shown in Table \ref{vlp_compare}, PixelBERT \cite{DBLP:journals/corr/abs-2004-00849/pixel-bert} only uses data in the field such as MS-COCO and VG, while VilBERT \cite{vilbert} and Unicoder-VL \cite{unicoder} only use CC. ViLT \cite{kim2021vilt} and UNITER \cite{uniter} pretrained on both in-domain (MS-COCO and VG) and out-of-domain (CC and SBU) datasets. It is easy to see from Table \ref{vlp_result} that as the amount of pre-training data scales up, the models get better retrieval results despite other factors such as model architecture and modality interaction. For example, ViLBERT vs. ViLT, have different model architectures, but ViLT with more pre-training data obtains better retrieval results in all metrics. In fact, Unicoder-VL and UNITER nearly belong to the same architecture, and the only difference is the pre-training datasets, so naturally, in most cases, UNITER gets better retrieval results, except TR@1 and TR@10 in Flickr30k, as the original paper reported.
PixelBERT-R50 and ViLT both use light visual tokens, such as CNN and direct image patches, and the same single cross-modal transformer. Clearly shown in Table \ref{vlp_result}, ViLT exceeds PixelBERT-R50 in every retrieval metric with a large margin due to the larger pre-training data though other factors changes.

This finding was also confirmed in the original ablation experiments of many papers, but this has provoked our concern. Even if the paper provides the original pretraining code, it will not be replicated owing to a lack of information and prohibitively high cost.
Even if researchers have the resources for reproduction, they are unwilling to devote too much money and resource consumption within a limited time and unexplained details~\cite{DBLP:conf/acl/BianchiH21}. Instead of reproducing these pretraining models, they could directly use the provided checkpoints in the pretraining stage. This manner is commonly preferred by small institutions, schools, and independent researchers. However, it is unknown whether the offered checkpoints required any pretraining abilities, included data from downstream tasks, or required extra manual annotation.

\subsubsection{Concerning of additional settings}\label{sec:ad}\ \\
\textbf{\textit{Improvement of these VLP models may not only come from the pretraining and architecture design but also have their own tricks and unknown details.}}

\textbf{Tricks.} The primary difference between the original paper and our reproduction in  ViLBERT and UNITER is the usage of online in-batch hard negative mining.
As shown in Table \ref{vlp_compare} and Table \ref{vlp_result}, we focus on the image-text retrieval task and use the checkpoints provided by the original paper to make a certain comparison. At this time, the training rounds of most models are the same as the warmup strategy. 
As \citet{DBLP:conf/emnlp/ZhangHJIS20} said, hard negative mining was added according to the description of ViLBERT's original paper, where a hard negative was selected from among the 100 closest neighbors of the target image by using the settings shown in Table \ref{vlp_compare}.
The results of the two datasets outperform those of the original paper, and several values that were not given in the original paper are included. 
For UNITER, the hard negative mining method provides the open-source code. However, after practice, we discover that this method is too time-consuming. In the original paper, the authors carried out the forward propagation of the network through the network model at a certain time, obtained $M$ negative samples, and then took the most difficult $N$ samples as the hard negative samples.
On MS-COCO, its $M$ setting is 399 and $N$ setting is 31. 
Even with 16 A100 GPUs, ignoring the time for backpropagation and sorting negative samples, the calculation of forward propagation on UNITER-base (111M) is approaching 125 hours.
It is impossible for researchers with limited resources to reproduce in a short time. Therefore, we did not reproduce the hard negative mining results of UNITER. Instead, by loading the released pretrained model of UNITER-Base and fine-tuning it on MS-COCO and Flickr30K, new results can be obtained, as shown in Table \ref{vlp_result}. This shows that without the hard sample mining operation, the reproducible results can make a huge difference, e.g. IR@1 and TR@1 drop by an average of 10 points on both Flickr30k and MS-COCO datasets. 
The only differences between UNITER-Base and ViLT are the visual-feature embedding and pretraining tasks. However, it can be seen that only a portion of the replicated UNITER-Base's results is close to ViLT, and most indicators, such as Flickr30k TR@1, TR@5, and TR@10, and the MS-COCO dataset, have a considerable reduction. Hard negative samples have a significant impact on the image-text retrieval model, as can be observed.

\textbf{Unknown details.} For PixelBERT and Unicode-VL, even if they have a large number of references and great influence, we still cannot obtain comparable results. On the one hand, due to the lack of pretrained checkpoints, we cannot obtain the results of the pretraining stage. On the other hand, due to the lack of details, it is also unknown how much influences the fuzziness of training rounds and hyperparametric settings, as well as the sampling method of hard samples and data enhancement. For PixelBERT, there is a lack of training details, such as hyperparameter settings in the pretraining stage and retrieval stage. Unicode-VL adopts the hard negative method of  \citet{DBLP:conf/iclr/RobinsonCSJ21/hnm}, but we are unable to duplicate it due to a lack of specifics.

The modal embedding and similarity modeling approaches are comparable in ViLBERT (221M) and UNITER-Base (110M), but the modality interaction, parameters, and amount of pretrained data are different. Although UNITER-Base uses more pretrained data, ViLBERT still outperforms UNITER-Base on numerous retrieval indicators (MS-COCO in R@1, 5, and 10) due to the combination of a larger parameter amount, modality interaction of the co-attention mechanism, and hard negative mining. We almost reported a number that was close to the original paper by loading the authors' pre-trained checkpoints, but the training took longer due to the uncertainty exacerbated by random data augmentation. 

\begin{figure*}[t]
    \centering
    \includegraphics[width=0.8\linewidth]{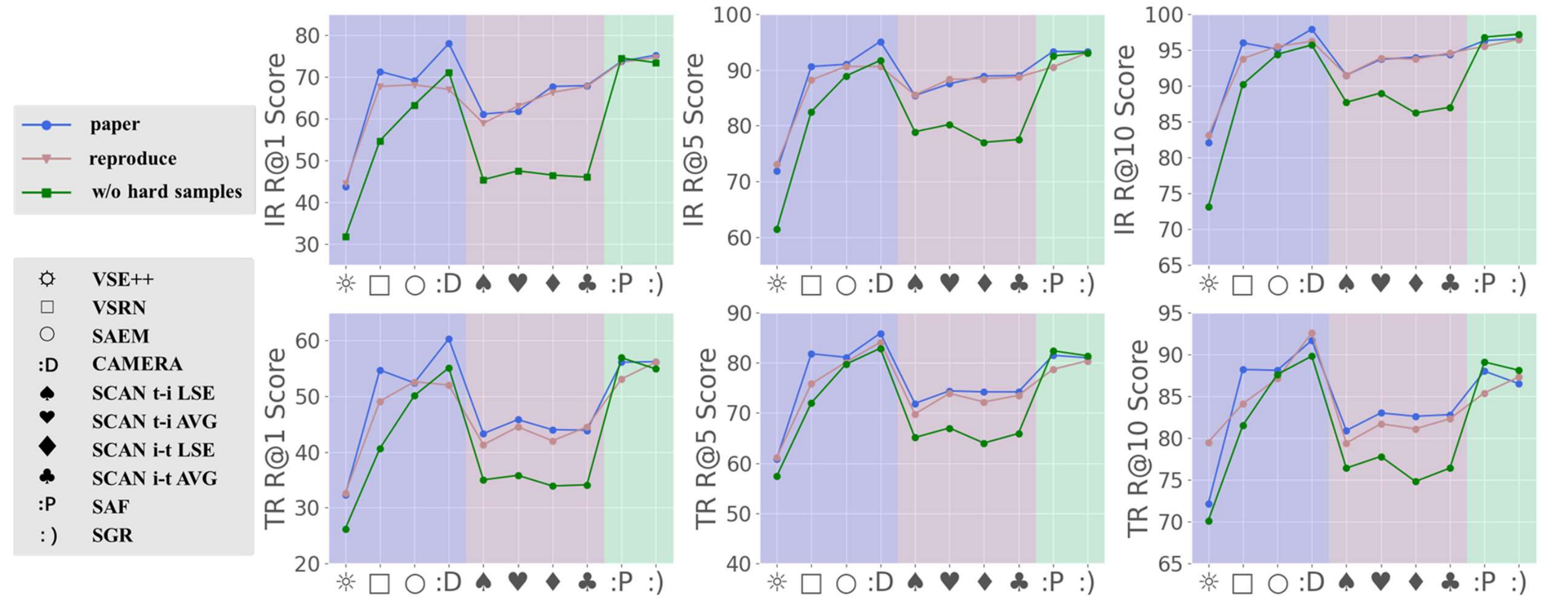}
    \caption{The comparison on Flickr30k dataset of \textcolor[RGB]{65,105,225}{the experimental results in original papers}, \textcolor[RGB]{195,142,145}{the reproduced results}, and \textcolor[RGB]{0,128,0}{the results of removing hard samples}. The different colored backgrounds correspond to different types of structures in Figure  \ref{fig:overview} \textcolor[RGB]{194,194,232}{(a)}, \textcolor[RGB]{216,201,216}{(b)} and \textcolor[RGB]{201,232,216}{(c)}, respectively. 
}
    \label{fig:f30k}
\end{figure*}

\begin{figure*}[t]
    \centering
    \includegraphics[width=0.8\linewidth]{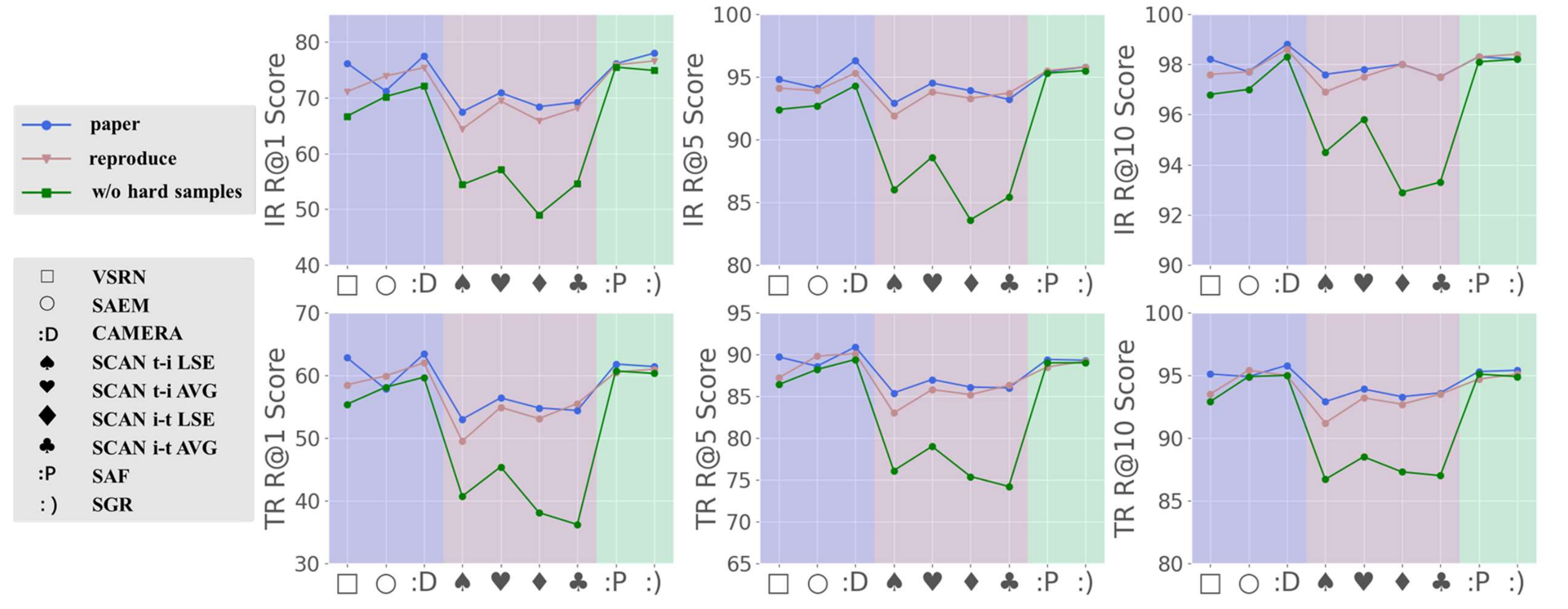}
    \caption{The comparison on MS-COCO (1K test) dataset of \textcolor[RGB]{65,105,225}{the experimental results in original papers}, \textcolor[RGB]{195,142,145}{the reproduced results}, and \textcolor[RGB]{0,128,0}{the results of removing hard samples}. 
    }
    \label{fig:coco1k}    
\end{figure*}

\begin{figure*}[t]
    \centering
    \includegraphics[width=0.8\linewidth]{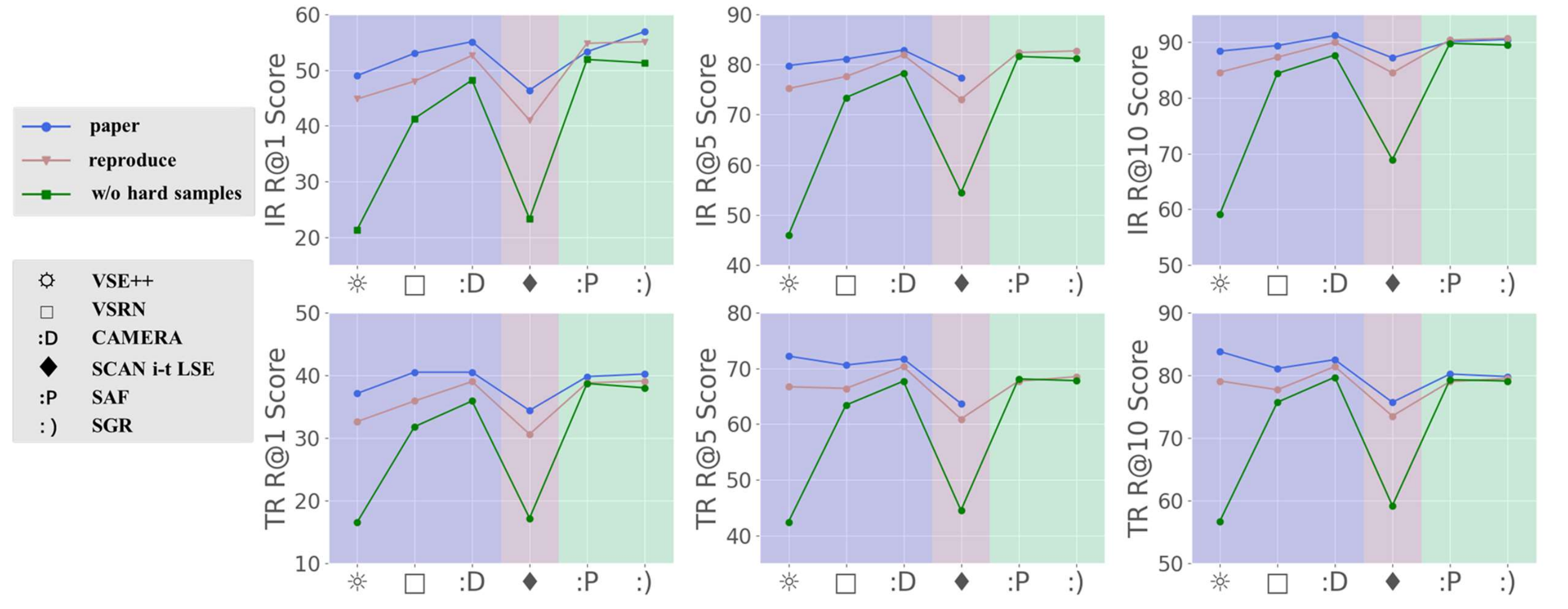}
    \caption{The comparison on MS-COCO (5K test) dataset of \textcolor[RGB]{65,105,225}{the experimental results in original papers}, \textcolor[RGB]{195,142,145}{the reproduced results}, and \textcolor[RGB]{0,128,0}{the results of removing hard samples}. 
    The paper proposed :P (SGR) and :) (SAF) has not provided R@5 result on MS-COCO (5K test) dataset.}
    \label{fig:coco5k}    
\end{figure*}

\subsection{NonPretrained Models}
\begin{figure*}[h]
    \centering
    \includegraphics[width=0.33\linewidth]{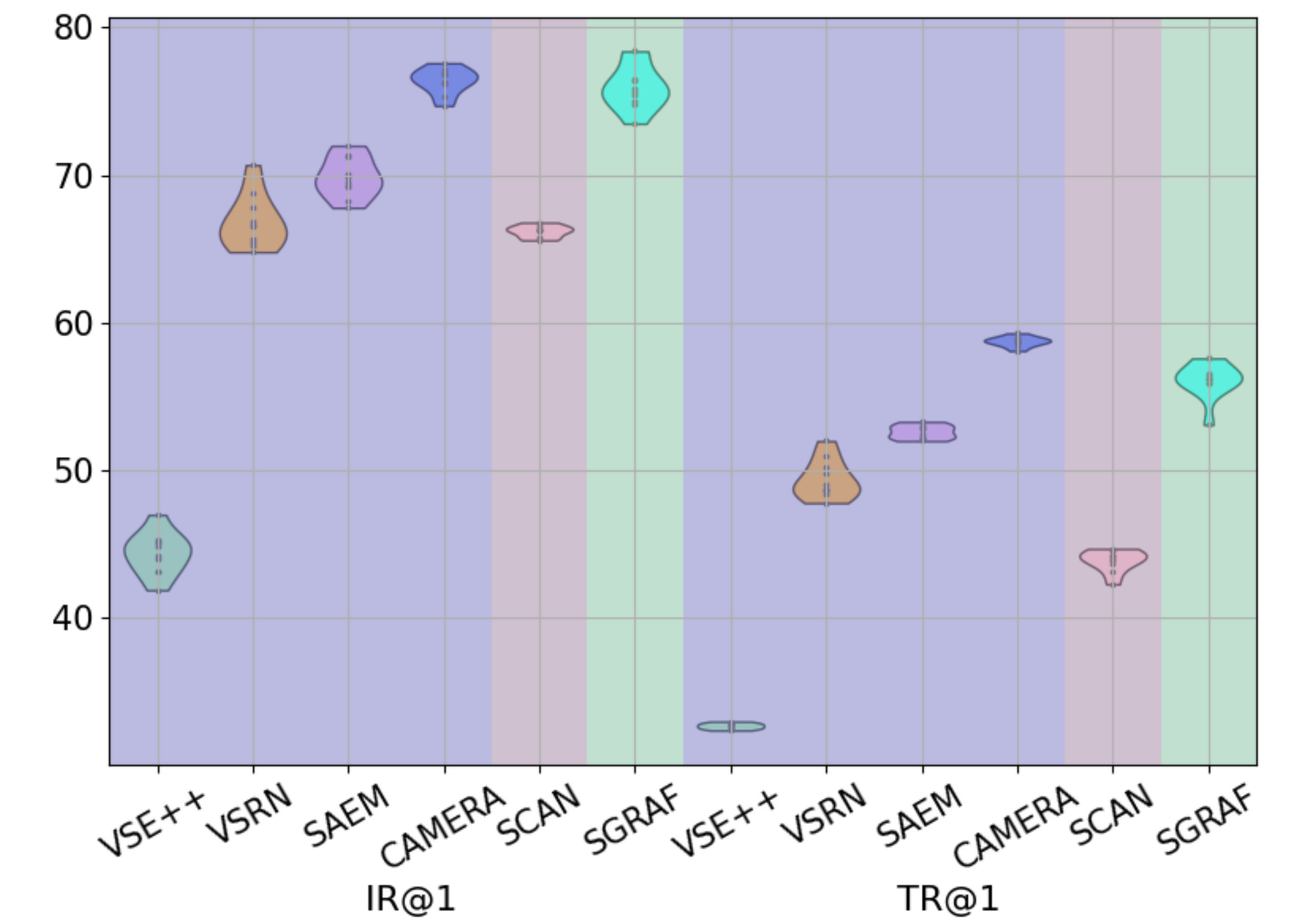}
    \includegraphics[width=0.33\linewidth]{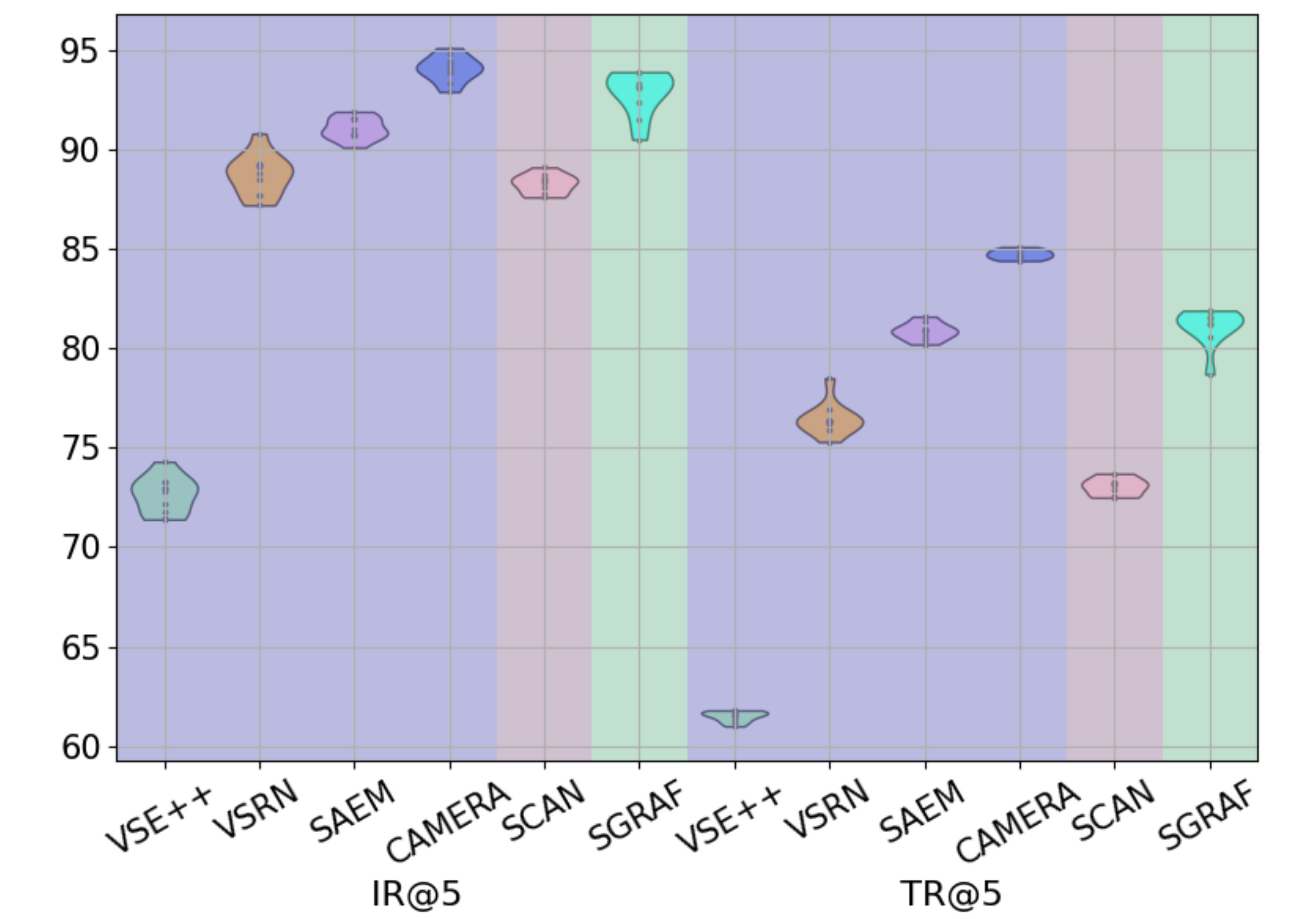}
    \includegraphics[width=0.33\linewidth]{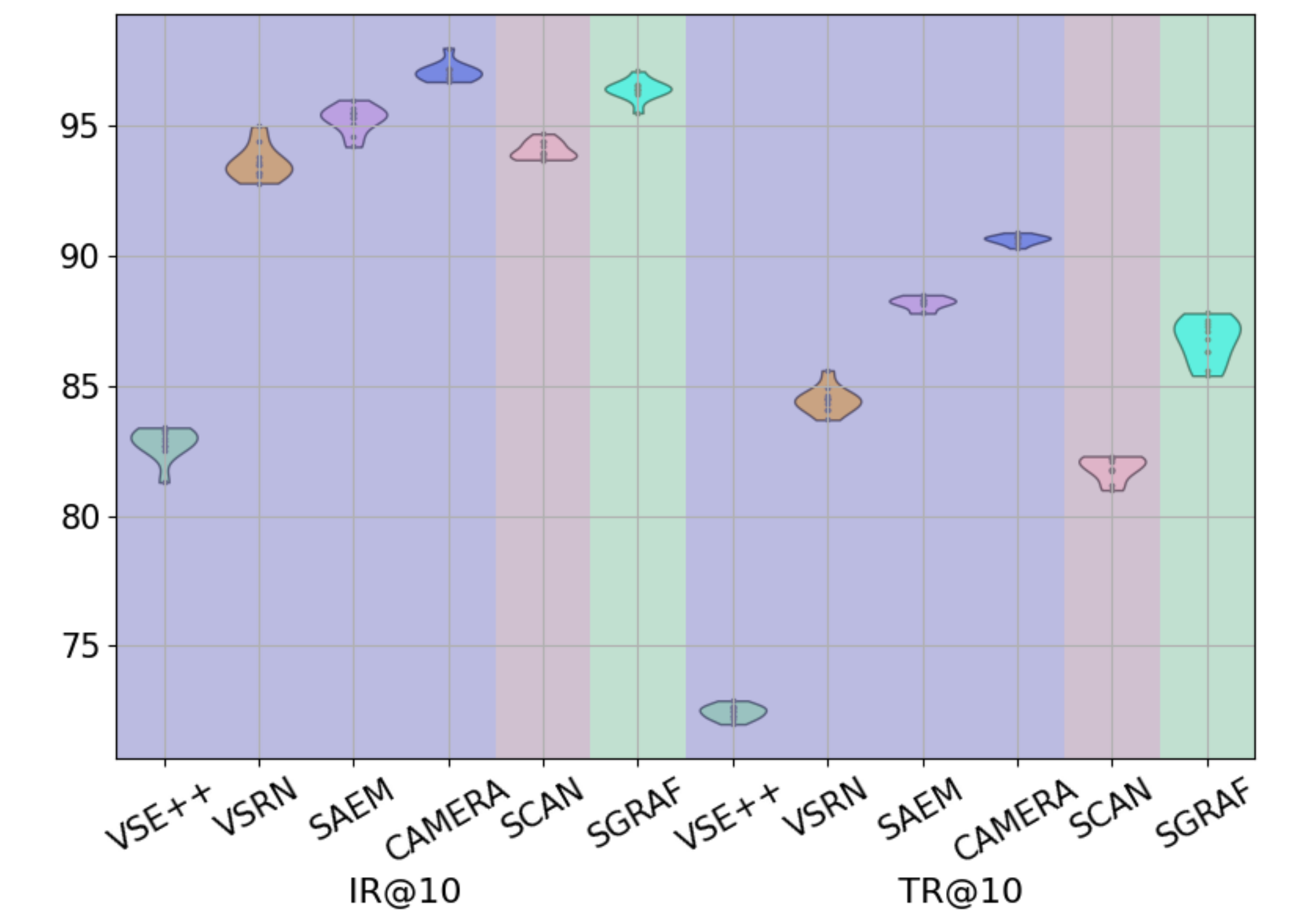}
    \caption{Fine-tuning variance in nonpretrained models on Flickr30k. Each model is fine-tuned 10 times with different random seeds. For SCAN/SGRAF, we chose the SCAN i-t ANG/SGR result to show.}
    \label{fig:f30k_avg}    
\end{figure*}

Because of lower calculation consumption and fewer parameters, the nonpretrained approach is also an essential component in driving the development of the image-text retrieval community and more ablation experiments can be carried out.
The above findings of VLPs lead us to consider whether the modality interaction style and the use of hard samples are also key factors in performance improvement. 
Therefore, in the following section, we discuss the relevant contents in nonpretrained models. We compare existing nonpretrained image-text retrieval models and present their detailed settings and parameter comparisons in Table \ref{tab:non_vlp_setup}. Nonpretrained models for image-text retrieval are more likely to create complicated modality interactions and get more effective results than pretrained models. After VSE++ \cite{vse++} is published, hard samples are used in nonpretrained methods widely. To research the roles of modality interaction and hard samples in image-text retrieval tasks, we reproduce several of existing nonpretrained image-text retrieval models with public codes and show results on Flickr30K dataset with Figure \ref{fig:f30k} and on MS-COCO dataset with Figure \ref{fig:coco1k}/Figure \ref{fig:coco5k}. We also show ten other experimental results with different random seeds on Flickr30k in Figure \ref{fig:f30k_avg}. We tried to find some interesting conclusions from the factors that affect nonpretrained models.

\subsubsection{Concerning of the environment and code.}\label{sec:non-pretrain_size}\ \\
\textbf{\textit{Reproducing the nonpretrained image-text retrieval models is not a trivial task.}}
The majority of the code in the papers we collected was written using the older torch framework version and Python 2. To run the code on Tesla A100 with torch 1.8.0 and CUDA 11, we just updated the code without changing the function of the program. We replicate each approach using the settings described in the original articles and the run statement in the README.md file in their codebase. 

SCAN \cite{SCAN}, SGRAF \cite{DBLP:conf/aaai/DiaoZML21/SGRAF}, VSRN \cite{DBLP:conf/iccv/LiZLLF19/VSRN}, CAMERA \cite{DBLP:conf/mm/Qu0CN020/campera}, and SAEM \cite{DBLP:conf/mm/WuWSH19/saem} provide accurate training/testing codes to reproduce relatively easily. 
While, SAEM \cite{DBLP:conf/mm/WuWSH19/saem} does not provide a test code on MS-COCO 1,000 test. 
Some of the methods do not provide the complete results on Flickr30k and MS-COCO.
VSE++ \cite{vse++} includes tricks on using the validation set for training, preprocessing images with a single random or center crop, and finetuning the image feature extractor.
We removed the above tricks to conduct our experiments.
Moreover, the data selection for the five-fold cross-validation of the MS-COCO 1,000 test was not random, which also reduces the credibility of all methods. Therefore, a unified code framework and reasonable testing methods are some of the important factors to promote the orderly development of the image-text retrieval community. 
In addition to the above, we discovered that few approaches were replicated in the papers gathered, and the results in the original publications were directly listed.
In such a manner, it is difficult for the community to know what lessons from previous research have held up, and it is tough for future researchers to improve on them. 



\subsubsection{Impact of different random seeds.}\label{sec:random_seeds}\ \\
\textbf{\textit{The raising of the metric score may come from different random seeds rather than the improvement of methods.}}
Stability is required for research that promotes the development of image-text retrieval. This means that future researchers can more easily reproduce and improve existing methods.
The stability of the method is usually manifested in the degree of dispersion of the results of multiple experiments \cite{lawrence2020almost}.
When we looked at the source code for some methods, we were surprised to find that even though they were available, it was hard to reproduce some of them with the same value. We show the results presented in the paper, our reproduced results, and the results after removing the hard sample method in Figure \ref{fig:f30k}, Figure \ref{fig:coco1k}, and Figure  \ref{fig:coco5k}, corresponding to the test results on Flickr30K dataset, 1,000 test set, and 5,000 test set of MS-COCO dataset, respectively.
As can be seen from the three figures, the reproduction results of CAMERA \cite{DBLP:conf/mm/Qu0CN020/campera} on the Flick30k dataset and VSE++ \cite{vse++}/VSRN \cite{DBLP:conf/iccv/LiZLLF19/VSRN} on the MS-COCO (1K test) dataset have a slight gap compared with their reported results.
To further investigate the stability and how this gap occurs, we reproduce them using more random seeds and show the results in Figure \ref{fig:f30k_avg}. As expected, the reproduced results of different methods fluctuated within a range. Surprisingly, the reproduced scores of the most volatile method can range  nearly 
10+ points. In the image-text retrieval task, the test setup in many papers is not mentioned, so many methods may pick the best one among many experimental results. This is unjust, as it fails to assess whether approaches are of higher quality. So more fair test methods, such as average and n-fold cross-validation, must be introduced.

\blfootnote{Figure \ref{fig:coco1k} and Figure \ref{fig:coco5k} lack a part of methods' result because they are not provided in papers or the corresponding code that can reproduce results is not provided. We provide more detailed result on https://github.com/WangFei-2019/Image-text-Retrieval.}

\subsubsection{Impact of multi-modal interactions.} \label{sec:modal_interaction}\ \\
\textbf{\textit{Modality interactions help improve the stability of models.}} According to the categories classified in Figure \ref{fig:overview}, we find that the 
(c) structures have a superior combination of modality encoding and modality alignment. The related approaches also have closer replicable results to the original papers' results even without hard samples, such as SGRAF \cite{DBLP:conf/aaai/DiaoZML21/SGRAF}. They also have better performance and a more aggregated distribution in Figure \ref{fig:f30k_avg}.
In VLP, we saw a similar conclusion as shown in Table \ref{vlp_result}. The single-stream VLP model is more capable of modality interaction than the two-stream VLP model. 
As a result, figuring out how to better align visual and text modalities remains a key breakthrough point for improving image-text retrieval performance. 
We found that the removing hard samples
results of SAEM \cite{DBLP:conf/mm/WuWSH19/saem} and CAMERA \cite{DBLP:conf/mm/Qu0CN020/campera} are near to the experimental results in the corresponding papers. The most notable distinction between SAEM \cite{DBLP:conf/mm/WuWSH19/saem}/CAMERA \cite{DBLP:conf/mm/Qu0CN020/campera} and other nonpretrained models is that the former two use the pretrained BERT as the text encoder, which has trained with a large number of the corpus. 

It is worth noting that the text encoder is trainable while the image encoder is non-trainable in image-text retrieval settings, which leads to unstable results when fewer training steps and random initialization are adopted.
Compared with LSTM and GRU, BERT is a more efficient text encoder and makes image-text retrieval methods more stable with self-supervised pretraining in large scale corpus. But this pretraining of the text encoder makes it hard to know whether the improvement comes from the text encoder or the network architecture.

\subsubsection{Impact of hard samples}\label{sec:non_hm}\ \\
\textbf{\textit{The use of hard samples or proper modality encoding and modality alignment contribute to the similarity modeling ability of models.}}
Using hard samples in image-text retrieval tasks is proposed by VSE++ \cite{vse++}, which is a method to assist similarity modeling. We have added experiments, shown in Figure \ref{fig:f30k}, Figure \ref{fig:coco1k} and Figure \ref{fig:coco5k}, to remove hard samples to verify the modeling ability of the model itself. \textit{The performance of almost all methods degrades to some extent after removing hard samples, except for the model with the structure in Figure \ref{fig:overview} (c)}.
This further reveals that the (c) in Figure \ref{fig:overview} has better similarity modeling ability. However, whether VSE++ \cite{vse++}, VSRN \cite{DBLP:conf/iccv/LiZLLF19/VSRN}, SAEM \cite{DBLP:conf/mm/WuWSH19/saem}, and CAMERA \cite{DBLP:conf/mm/Qu0CN020/campera} which have an excellent modality encoding ability, or SCAN \cite{SCAN} which has a superior modality alignment ability, all get a bad performance without hard samples. Therefore, in addition to using hard samples, boosting the model's modality encoding and modality alignment abilities are essential ways to improve similarity modeling ability.



\section{Conclusion and suggestion}

As discussed above, we were astonished that so few of the architectural alterations of modality interaction and similarity modeling resulted in gains, even when we used nearly the same parameters as the original paper but omitted some techniques and used different random seeds.
There are several probable explanations for why our findings were as they were:\\%
\textit{1. Training data scale is the key. (\S \ref{sec:data_size} and \S \ref{sec:non-pretrain_size})}
In the VLP results, we found that the addition of data can significantly improve the final results but has worse reproducibility. Meanwhile, when we reproduced the nonpretrained model, the original text of the specific operation of data enhancement and the description of the code is too vague, so the data comparison is not necessarily carried out under the exact same settings. Moreover, we found that for the larger dataset MS-COCO and the harder 5K test, the results of each method run are more stable. In the process of reproduction, we did not use the validation set data, so this quantitative data may have caused a certain degree of decline in the reproduction results.\\
\textit{2. Additional settings are vital. (\S \ref{sec:ad})} We discovered that, despite being thought to be some ``tricks'', some of the details omitted by the authors from the paper played a significant role. We wish that the authors could have described them in detail to help the researchers really understand where the enhancements came from.
\\
\textit{3. Not tuning hyperparameters handicapped other methods. (\S \ref{sec:random_seeds})} In our replication, we discovered that the random seed has a significant impact on the experimental results, with differences of up to nearly 10 points on IR@1 (see Figure \ref{fig:f30k_avg}). As a result, these modification strategies are not sufficiently hyperparameter-agnostic and stable in their modality interaction modification. Furthermore, if parameter sensitivity is the key to the problem, using a decent initialization as the final result does not reflect one's own method's contribution. 
\\
\textit{4. Modality interaction types may be critical. (\S \ref{sec:modal_interaction})}
For many VLPs, we can use most of their methods as simple transformers as modality interactions. Although the co-attention mechanism of ViLBERT has a marginal performance improvement, it introduces double the parameters, resulting in greater computational consumption.
For many nonpretrained models, most put the direction of improvement on modality interaction. To our surprise, in addition to improving the performance of the model to a certain extent, the multi-modal interaction of type (c) and multi-modal feature (type (d)) can also make the model more stable(see Figure \ref{fig:f30k}, \ref{fig:coco1k}, and \ref{fig:coco5k}).  
\\
\textit{5. How to make better use of training samples is also the key. (\S \ref{sec:ad} and \S \ref{sec:non_hm})}
Hard samples are widely used in both VLPs and nonpretrained models. Through ablation experiments on hard negative mining in VLPs experiments, we found that more hard samples can greatly enhance the retrieval results. In the nonpretrained results, we found that the use of the most hard samples was not the same as that of VLPs. These methods only use the most similar negative samples and one positive sample in a batch for optimization, resulting in less utilization of the characteristics of the data. Additionally, due to the parameter sensitivity of these methods, under different random seeds, the model similarity modeling ability is different. This, in turn, magnifies the effect of hard samples. 

Given these findings, we propose some suggestions to improve the robustness and generalizability of future image-text retrieval research. 
First, when proposing a new method, the random seed used and the results of multiple runs and specific details should be given. The best-practice results reporting should include the \textbf{mean and standard deviation} across numerous trials or at the very least avoid cherry-picking the best run \cite{DBLP:conf/emnlp/NarangCTFFMMFSL21} like Figure  \ref{fig:f30k_avg}.
Second, we should not pay too much attention to the tuning of models and parameters, and we should focus on the characteristics of the mining \textbf{data}. Hard samples can bring a more stable improvement to the model \cite{DBLP:conf/iclr/RobinsonCSJ21/hnm}, but such methods are rarely used in the field of image-text retrieval. Based on the findings of this recurring result, we believe that in the future, we should focus on such methods of stable improvement rather than those that need to be sensitive to parameters.
Finally, we should rigorously evaluate models using \textbf{more than one metric} to get a comprehensive understanding of how a good method works in different situations,
such as NDCG~\cite{ndcg} and mAP~\cite{map}. In a realistic scene, users are searching for relevant images/captions but not necessarily exact matches. The Recall@K evaluation provides results that are too rigid, which ignores other relevant but not exact-matching elements that users may be interested in.


\begin{acks}
This research was funded by Major Scientific and Technological Projects of CNPC (Grant No.ZD2019-183-008), the National Natural Science Foundation of China (Grant No.61671480), and the Open Program of the National Laboratory of Pattern Recognition (Grant No.202000009), National Natural Science Foundation of China (No.61902093), Natural Science Foundation of Guangdong (No.2020A1515010652), Shenzhen Foundational Research Funding Under Grant (No.20200805173048001), and PINGAN-HITsz Intelligence Finance Research Center.
\end{acks}
\bibliographystyle{ACM-Reference-Format}
\bibliography{sample-sigconf}


\begin{thebibliography}{59}


\ifx \showCODEN    \undefined \def \showCODEN     #1{\unskip}     \fi
\ifx \showDOI      \undefined \def \showDOI       #1{#1}\fi
\ifx \showISBNx    \undefined \def \showISBNx     #1{\unskip}     \fi
\ifx \showISBNxiii \undefined \def \showISBNxiii  #1{\unskip}     \fi
\ifx \showISSN     \undefined \def \showISSN      #1{\unskip}     \fi
\ifx \showLCCN     \undefined \def \showLCCN      #1{\unskip}     \fi
\ifx \shownote     \undefined \def \shownote      #1{#1}          \fi
\ifx \showarticletitle \undefined \def \showarticletitle #1{#1}   \fi
\ifx \showURL      \undefined \def \showURL       {\relax}        \fi
\providecommand\bibfield[2]{#2}
\providecommand\bibinfo[2]{#2}
\providecommand\natexlab[1]{#1}
\providecommand\showeprint[2][]{arXiv:#2}

\bibitem[Anderson et~al\mbox{.}(2018)]%
        {DBLP:conf/cvpr/00010BT0GZ18/bottom-up-attention}
\bibfield{author}{\bibinfo{person}{Peter Anderson}, \bibinfo{person}{Xiaodong
  He}, \bibinfo{person}{Chris Buehler}, \bibinfo{person}{Damien Teney},
  \bibinfo{person}{Mark Johnson}, \bibinfo{person}{Stephen Gould}, {and}
  \bibinfo{person}{Lei Zhang}.} \bibinfo{year}{2018}\natexlab{}.
\newblock \showarticletitle{Bottom-Up and Top-Down Attention for Image
  Captioning and Visual Question Answering}. In
  \bibinfo{booktitle}{\emph{CVPR}}.
\newblock


\bibitem[Bahdanau et~al\mbox{.}(2015)]%
        {DBLP:journals/corr/BahdanauCB14/gru}
\bibfield{author}{\bibinfo{person}{Dzmitry Bahdanau},
  \bibinfo{person}{Kyunghyun Cho}, {and} \bibinfo{person}{Yoshua Bengio}.}
  \bibinfo{year}{2015}\natexlab{}.
\newblock \showarticletitle{Neural Machine Translation by Jointly Learning to
  Align and Translate}. In \bibinfo{booktitle}{\emph{ICLR}}.
\newblock


\bibitem[Bianchi and Hovy(2021)]%
        {DBLP:conf/acl/BianchiH21}
\bibfield{author}{\bibinfo{person}{Federico Bianchi} {and}
  \bibinfo{person}{Dirk Hovy}.} \bibinfo{year}{2021}\natexlab{}.
\newblock \showarticletitle{On the Gap between Adoption and Understanding in
  {NLP}}. In \bibinfo{booktitle}{\emph{Findings of ACL}}.
\newblock


\bibitem[Chen et~al\mbox{.}(2020)]%
        {uniter}
\bibfield{author}{\bibinfo{person}{Yen{-}Chun Chen}, \bibinfo{person}{Linjie
  Li}, \bibinfo{person}{Licheng Yu}, \bibinfo{person}{Ahmed~El Kholy},
  \bibinfo{person}{Faisal Ahmed}, \bibinfo{person}{Zhe Gan},
  \bibinfo{person}{Yu Cheng}, {and} \bibinfo{person}{Jingjing Liu}.}
  \bibinfo{year}{2020}\natexlab{}.
\newblock \showarticletitle{{UNITER:} UNiversal Image-TExt Representation
  Learning}. In \bibinfo{booktitle}{\emph{ECCV}}.
\newblock


\bibitem[Conneau and Lample(2019)]%
        {conneau2019cross}
\bibfield{author}{\bibinfo{person}{Alexis Conneau} {and}
  \bibinfo{person}{Guillaume Lample}.} \bibinfo{year}{2019}\natexlab{}.
\newblock \showarticletitle{Cross-lingual language model pretraining}.
\newblock \bibinfo{journal}{\emph{NeurIPS}} (\bibinfo{year}{2019}).
\newblock


\bibitem[Devlin et~al\mbox{.}(2019)]%
        {bert}
\bibfield{author}{\bibinfo{person}{Jacob Devlin}, \bibinfo{person}{Ming{-}Wei
  Chang}, \bibinfo{person}{Kenton Lee}, {and} \bibinfo{person}{Kristina
  Toutanova}.} \bibinfo{year}{2019}\natexlab{}.
\newblock \showarticletitle{{BERT:} Pre-training of Deep Bidirectional
  Transformers for Language Understanding}. In
  \bibinfo{booktitle}{\emph{NAACL-HLT}}. \bibinfo{publisher}{Association for
  Computational Linguistics}.
\newblock


\bibitem[Diao et~al\mbox{.}(2021)]%
        {DBLP:conf/aaai/DiaoZML21/SGRAF}
\bibfield{author}{\bibinfo{person}{Haiwen Diao}, \bibinfo{person}{Ying Zhang},
  \bibinfo{person}{Lin Ma}, {and} \bibinfo{person}{Huchuan Lu}.}
  \bibinfo{year}{2021}\natexlab{}.
\newblock \showarticletitle{Similarity Reasoning and Filtration for Image-Text
  Matching}. In \bibinfo{booktitle}{\emph{AAAI}}.
\newblock


\bibitem[Ding et~al\mbox{.}(2021)]%
        {ding2021progressive}
\bibfield{author}{\bibinfo{person}{Liang Ding}, \bibinfo{person}{Longyue Wang},
  \bibinfo{person}{Xuebo Liu}, \bibinfo{person}{Derek~F Wong},
  \bibinfo{person}{Dacheng Tao}, {and} \bibinfo{person}{Zhaopeng Tu}.}
  \bibinfo{year}{2021}\natexlab{}.
\newblock \showarticletitle{Progressive Multi-Granularity Training for
  Non-Autoregressive Translation}. In \bibinfo{booktitle}{\emph{Fingdings of
  ACL}}.
\newblock


\bibitem[Ding et~al\mbox{.}(2020)]%
        {ding2020self}
\bibfield{author}{\bibinfo{person}{Liang Ding}, \bibinfo{person}{Longyue Wang},
  {and} \bibinfo{person}{Dacheng Tao}.} \bibinfo{year}{2020}\natexlab{}.
\newblock \showarticletitle{Self-attention with cross-lingual position
  representation}. In \bibinfo{booktitle}{\emph{ACL}}.
\newblock


\bibitem[Dosovitskiy et~al\mbox{.}(2020)]%
        {dosovitskiy2020image}
\bibfield{author}{\bibinfo{person}{Alexey Dosovitskiy}, \bibinfo{person}{Lucas
  Beyer}, \bibinfo{person}{Alexander Kolesnikov}, \bibinfo{person}{Dirk
  Weissenborn}, \bibinfo{person}{Xiaohua Zhai}, \bibinfo{person}{Thomas
  Unterthiner}, \bibinfo{person}{Mostafa Dehghani}, \bibinfo{person}{Matthias
  Minderer}, \bibinfo{person}{Georg Heigold}, \bibinfo{person}{Sylvain Gelly},
  {et~al\mbox{.}}} \bibinfo{year}{2020}\natexlab{}.
\newblock \showarticletitle{An image is worth 16x16 words: Transformers for
  image recognition at scale}. In \bibinfo{booktitle}{\emph{ICLR}}.
\newblock


\bibitem[Faghri et~al\mbox{.}(2018)]%
        {vse++}
\bibfield{author}{\bibinfo{person}{Fartash Faghri}, \bibinfo{person}{David~J.
  Fleet}, \bibinfo{person}{Jamie~Ryan Kiros}, {and} \bibinfo{person}{Sanja
  Fidler}.} \bibinfo{year}{2018}\natexlab{}.
\newblock \showarticletitle{{VSE++:} Improving Visual-Semantic Embeddings with
  Hard Negatives}. In \bibinfo{booktitle}{\emph{BMVC}}.
\newblock


\bibitem[Gao et~al\mbox{.}(2020)]%
        {DBLP:conf/sigir/GaoJCQLWHW20/fashionbert}
\bibfield{author}{\bibinfo{person}{Dehong Gao}, \bibinfo{person}{Linbo Jin},
  \bibinfo{person}{Ben Chen}, \bibinfo{person}{Minghui Qiu},
  \bibinfo{person}{Peng Li}, \bibinfo{person}{Yi Wei}, \bibinfo{person}{Yi Hu},
  {and} \bibinfo{person}{Hao Wang}.} \bibinfo{year}{2020}\natexlab{}.
\newblock \showarticletitle{FashionBERT: Text and Image Matching with Adaptive
  Loss for Cross-modal Retrieval}. In \bibinfo{booktitle}{\emph{SIGIR}}.
\newblock


\bibitem[He et~al\mbox{.}(2016)]%
        {DBLP:conf/cvpr/HeZRS16/resnet}
\bibfield{author}{\bibinfo{person}{Kaiming He}, \bibinfo{person}{Xiangyu
  Zhang}, \bibinfo{person}{Shaoqing Ren}, {and} \bibinfo{person}{Jian Sun}.}
  \bibinfo{year}{2016}\natexlab{}.
\newblock \showarticletitle{Deep Residual Learning for Image Recognition}. In
  \bibinfo{booktitle}{\emph{CVPR}}.
\newblock


\bibitem[Hu et~al\mbox{.}(2019b)]%
        {DBLP:conf/sigir/HuZPL19}
\bibfield{author}{\bibinfo{person}{Peng Hu}, \bibinfo{person}{Liangli Zhen},
  \bibinfo{person}{Dezhong Peng}, {and} \bibinfo{person}{Pei Liu}.}
  \bibinfo{year}{2019}\natexlab{b}.
\newblock \showarticletitle{Scalable Deep Multimodal Learning for Cross-Modal
  Retrieval}. In \bibinfo{booktitle}{\emph{SIGIR}}.
\newblock


\bibitem[Hu et~al\mbox{.}(2019a)]%
        {DBLP:conf/ijcai/HuLLYC19}
\bibfield{author}{\bibinfo{person}{Zhibin Hu}, \bibinfo{person}{Yongsheng Luo},
  \bibinfo{person}{Jiong Lin}, \bibinfo{person}{Yan Yan}, {and}
  \bibinfo{person}{Jian Chen}.} \bibinfo{year}{2019}\natexlab{a}.
\newblock \showarticletitle{Multi-Level Visual-Semantic Alignments with
  Relation-Wise Dual Attention Network for Image and Text Matching}. In
  \bibinfo{booktitle}{\emph{{IJCAI}}}.
\newblock


\bibitem[Huang et~al\mbox{.}(2020)]%
        {DBLP:journals/corr/abs-2004-00849/pixel-bert}
\bibfield{author}{\bibinfo{person}{Zhicheng Huang}, \bibinfo{person}{Zhaoyang
  Zeng}, \bibinfo{person}{Bei Liu}, \bibinfo{person}{Dongmei Fu}, {and}
  \bibinfo{person}{Jianlong Fu}.} \bibinfo{year}{2020}\natexlab{}.
\newblock \showarticletitle{Pixel-BERT: Aligning Image Pixels with Text by Deep
  Multi-Modal Transformers}.
\newblock \bibinfo{journal}{\emph{CoRR}}  \bibinfo{volume}{abs/2004.00849}
  (\bibinfo{year}{2020}).
\newblock


\bibitem[J{\"{a}}rvelin and Kek{\"{a}}l{\"{a}}inen(2002)]%
        {ndcg}
\bibfield{author}{\bibinfo{person}{Kalervo J{\"{a}}rvelin} {and}
  \bibinfo{person}{Jaana Kek{\"{a}}l{\"{a}}inen}.}
  \bibinfo{year}{2002}\natexlab{}.
\newblock \showarticletitle{Cumulated gain-based evaluation of {IR}
  techniques}.
\newblock \bibinfo{journal}{\emph{{ACM} Trans. Inf. Syst.}}
  (\bibinfo{year}{2002}).
\newblock


\bibitem[Jia et~al\mbox{.}(2021)]%
        {DBLP:conf/icml/JiaYXCPPLSLD21/scaling}
\bibfield{author}{\bibinfo{person}{Chao Jia}, \bibinfo{person}{Yinfei Yang},
  \bibinfo{person}{Ye Xia}, \bibinfo{person}{Yi{-}Ting Chen},
  \bibinfo{person}{Zarana Parekh}, \bibinfo{person}{Hieu Pham},
  \bibinfo{person}{Quoc~V. Le}, \bibinfo{person}{Yun{-}Hsuan Sung},
  \bibinfo{person}{Zhen Li}, {and} \bibinfo{person}{Tom Duerig}.}
  \bibinfo{year}{2021}\natexlab{}.
\newblock \showarticletitle{Scaling Up Visual and Vision-Language
  Representation Learning With Noisy Text Supervision}. In
  \bibinfo{booktitle}{\emph{ICML}}.
\newblock


\bibitem[Jiang et~al\mbox{.}(2020)]%
        {DBLP:conf/cvpr/JiangMRLC20}
\bibfield{author}{\bibinfo{person}{Huaizu Jiang}, \bibinfo{person}{Ishan
  Misra}, \bibinfo{person}{Marcus Rohrbach}, \bibinfo{person}{Erik~G.
  Learned{-}Miller}, {and} \bibinfo{person}{Xinlei Chen}.}
  \bibinfo{year}{2020}\natexlab{}.
\newblock \showarticletitle{In Defense of Grid Features for Visual Question
  Answering}. In \bibinfo{booktitle}{\emph{CVPR}}.
\newblock


\bibitem[Karpathy and Fei{-}Fei(2017)]%
        {DBLP:journals/pami/KarpathyF17}
\bibfield{author}{\bibinfo{person}{Andrej Karpathy} {and} \bibinfo{person}{Li
  Fei{-}Fei}.} \bibinfo{year}{2017}\natexlab{}.
\newblock \showarticletitle{Deep Visual-Semantic Alignments for Generating
  Image Descriptions}.
\newblock \bibinfo{journal}{\emph{TPAMI}} \bibinfo{volume}{39},
  \bibinfo{number}{4} (\bibinfo{year}{2017}), \bibinfo{pages}{664--676}.
\newblock


\bibitem[Kim et~al\mbox{.}(2021)]%
        {kim2021vilt}
\bibfield{author}{\bibinfo{person}{Wonjae Kim}, \bibinfo{person}{Bokyung Son},
  {and} \bibinfo{person}{Ildoo Kim}.} \bibinfo{year}{2021}\natexlab{}.
\newblock \showarticletitle{ViLT: Vision-and-Language Transformer Without
  Convolution or Region Supervision}. In \bibinfo{booktitle}{\emph{ICML}}.
\newblock


\bibitem[Lawrence et~al\mbox{.}(2020)]%
        {lawrence2020almost}
\bibfield{author}{\bibinfo{person}{Nathan Lawrence}, \bibinfo{person}{Philip
  Loewen}, \bibinfo{person}{Michael Forbes}, \bibinfo{person}{Johan Backstrom},
  {and} \bibinfo{person}{Bhushan Gopaluni}.} \bibinfo{year}{2020}\natexlab{}.
\newblock \showarticletitle{Almost Surely Stable Deep Dynamics}.
\newblock \bibinfo{journal}{\emph{NeurIPS}} (\bibinfo{year}{2020}).
\newblock


\bibitem[Lee et~al\mbox{.}(2018)]%
        {SCAN}
\bibfield{author}{\bibinfo{person}{Kuang-Huei Lee}, \bibinfo{person}{Xi Chen},
  \bibinfo{person}{Gang Hua}, \bibinfo{person}{Houdong Hu}, {and}
  \bibinfo{person}{Xiaodong He}.} \bibinfo{year}{2018}\natexlab{}.
\newblock \showarticletitle{Stacked cross attention for image-text matching}.
  In \bibinfo{booktitle}{\emph{ECCV}}.
\newblock


\bibitem[Li et~al\mbox{.}(2020a)]%
        {unicoder}
\bibfield{author}{\bibinfo{person}{Gen Li}, \bibinfo{person}{Nan Duan},
  \bibinfo{person}{Yuejian Fang}, \bibinfo{person}{Ming Gong}, {and}
  \bibinfo{person}{Daxin Jiang}.} \bibinfo{year}{2020}\natexlab{a}.
\newblock \showarticletitle{Unicoder-VL: {A} Universal Encoder for Vision and
  Language by Cross-Modal Pre-Training}. In \bibinfo{booktitle}{\emph{AAAI}}.
\newblock


\bibitem[Li et~al\mbox{.}(2019b)]%
        {DBLP:conf/iccv/LiZLLF19/VSRN}
\bibfield{author}{\bibinfo{person}{Kunpeng Li}, \bibinfo{person}{Yulun Zhang},
  \bibinfo{person}{Kai Li}, \bibinfo{person}{Yuanyuan Li}, {and}
  \bibinfo{person}{Yun Fu}.} \bibinfo{year}{2019}\natexlab{b}.
\newblock \showarticletitle{Visual Semantic Reasoning for Image-Text Matching}.
  In \bibinfo{booktitle}{\emph{ICCV}}.
\newblock


\bibitem[Li et~al\mbox{.}(2019a)]%
        {DBLP:journals/corr/abs-1908-03557/visualbert}
\bibfield{author}{\bibinfo{person}{Liunian~Harold Li}, \bibinfo{person}{Mark
  Yatskar}, \bibinfo{person}{Da Yin}, \bibinfo{person}{Cho{-}Jui Hsieh}, {and}
  \bibinfo{person}{Kai{-}Wei Chang}.} \bibinfo{year}{2019}\natexlab{a}.
\newblock \showarticletitle{VisualBERT: {A} Simple and Performant Baseline for
  Vision and Language}.
\newblock \bibinfo{journal}{\emph{CoRR}}  \bibinfo{volume}{abs/1908.03557}
  (\bibinfo{year}{2019}).
\newblock


\bibitem[Li et~al\mbox{.}(2021)]%
        {li2020unimo}
\bibfield{author}{\bibinfo{person}{Wei Li}, \bibinfo{person}{Can Gao},
  \bibinfo{person}{Guocheng Niu}, \bibinfo{person}{Xinyan Xiao},
  \bibinfo{person}{Hao Liu}, \bibinfo{person}{Jiachen Liu},
  \bibinfo{person}{Hua Wu}, {and} \bibinfo{person}{Haifeng Wang}.}
  \bibinfo{year}{2021}\natexlab{}.
\newblock \showarticletitle{{UNIMO:} Towards Unified-Modal Understanding and
  Generation via Cross-Modal Contrastive Learning}. In
  \bibinfo{booktitle}{\emph{ACL/IJCNLP}}.
\newblock


\bibitem[Li et~al\mbox{.}(2020b)]%
        {oscar}
\bibfield{author}{\bibinfo{person}{Xiujun Li}, \bibinfo{person}{Xi Yin},
  \bibinfo{person}{Chunyuan Li}, \bibinfo{person}{Pengchuan Zhang},
  \bibinfo{person}{Xiaowei Hu}, \bibinfo{person}{Lei Zhang},
  \bibinfo{person}{Lijuan Wang}, \bibinfo{person}{Houdong Hu},
  \bibinfo{person}{Li Dong}, \bibinfo{person}{Furu Wei}, \bibinfo{person}{Yejin
  Choi}, {and} \bibinfo{person}{Jianfeng Gao}.}
  \bibinfo{year}{2020}\natexlab{b}.
\newblock \showarticletitle{Oscar: Object-Semantics Aligned Pre-training for
  Vision-Language Tasks}. In \bibinfo{booktitle}{\emph{ECCV}}.
\newblock


\bibitem[Lin et~al\mbox{.}(2017)]%
        {DBLP:conf/cvpr/LinDGHHB17/fpn}
\bibfield{author}{\bibinfo{person}{Tsung{-}Yi Lin}, \bibinfo{person}{Piotr
  Doll{\'{a}}r}, \bibinfo{person}{Ross~B. Girshick}, \bibinfo{person}{Kaiming
  He}, \bibinfo{person}{Bharath Hariharan}, {and} \bibinfo{person}{Serge~J.
  Belongie}.} \bibinfo{year}{2017}\natexlab{}.
\newblock \showarticletitle{Feature Pyramid Networks for Object Detection}. In
  \bibinfo{booktitle}{\emph{CVPR}}. \bibinfo{pages}{936--944}.
\newblock


\bibitem[Lin et~al\mbox{.}(2014)]%
        {DBLP:conf/eccv/LinMBHPRDZ14}
\bibfield{author}{\bibinfo{person}{Tsung{-}Yi Lin}, \bibinfo{person}{Michael
  Maire}, \bibinfo{person}{Serge~J. Belongie}, \bibinfo{person}{James Hays},
  \bibinfo{person}{Pietro Perona}, \bibinfo{person}{Deva Ramanan},
  \bibinfo{person}{Piotr Doll{\'{a}}r}, {and} \bibinfo{person}{C.~Lawrence
  Zitnick}.} \bibinfo{year}{2014}\natexlab{}.
\newblock \showarticletitle{Microsoft {COCO:} Common Objects in Context}. In
  \bibinfo{booktitle}{\emph{ECCV}}.
\newblock


\bibitem[Lu et~al\mbox{.}(2019)]%
        {vilbert}
\bibfield{author}{\bibinfo{person}{Jiasen Lu}, \bibinfo{person}{Dhruv Batra},
  \bibinfo{person}{Devi Parikh}, {and} \bibinfo{person}{Stefan Lee}.}
  \bibinfo{year}{2019}\natexlab{}.
\newblock \showarticletitle{ViLBERT: Pretraining Task-Agnostic Visiolinguistic
  Representations for Vision-and-Language Tasks}. In
  \bibinfo{booktitle}{\emph{NeurIPS}}.
\newblock


\bibitem[Lu et~al\mbox{.}(2021)]%
        {DBLP:conf/acl/LuZL20}
\bibfield{author}{\bibinfo{person}{Xiaopeng Lu}, \bibinfo{person}{Tiancheng
  Zhao}, {and} \bibinfo{person}{Kyusong Lee}.} \bibinfo{year}{2021}\natexlab{}.
\newblock \showarticletitle{VisualSparta: An Embarrassingly Simple Approach to
  Large-scale Text-to-Image Search with Weighted Bag-of-words}. In
  \bibinfo{booktitle}{\emph{{ACL}}}.
\newblock


\bibitem[Narang et~al\mbox{.}(2021)]%
        {DBLP:conf/emnlp/NarangCTFFMMFSL21}
\bibfield{author}{\bibinfo{person}{Sharan Narang}, \bibinfo{person}{Hyung~Won
  Chung}, \bibinfo{person}{Yi Tay}, \bibinfo{person}{Liam Fedus},
  \bibinfo{person}{Thibault F{\'{e}}vry}, \bibinfo{person}{Michael Matena},
  \bibinfo{person}{Karishma Malkan}, \bibinfo{person}{Noah Fiedel},
  \bibinfo{person}{Noam Shazeer}, \bibinfo{person}{Zhenzhong Lan},
  \bibinfo{person}{Yanqi Zhou}, \bibinfo{person}{Wei Li}, \bibinfo{person}{Nan
  Ding}, \bibinfo{person}{Jake Marcus}, \bibinfo{person}{Adam Roberts}, {and}
  \bibinfo{person}{Colin Raffel}.} \bibinfo{year}{2021}\natexlab{}.
\newblock \showarticletitle{Do Transformer Modifications Transfer Across
  Implementations and Applications?}. In \bibinfo{booktitle}{\emph{EMNLP}}.
\newblock


\bibitem[Ordonez et~al\mbox{.}(2011)]%
        {DBLP:conf/nips/OrdonezKB11/sbu}
\bibfield{author}{\bibinfo{person}{Vicente Ordonez}, \bibinfo{person}{Girish
  Kulkarni}, {and} \bibinfo{person}{Tamara~L. Berg}.}
  \bibinfo{year}{2011}\natexlab{}.
\newblock \showarticletitle{Im2Text: Describing Images Using 1 Million
  Captioned Photographs}. In \bibinfo{booktitle}{\emph{NeurIPS}}.
\newblock


\bibitem[Philbin et~al\mbox{.}(2007)]%
        {map}
\bibfield{author}{\bibinfo{person}{James Philbin}, \bibinfo{person}{Ondrej
  Chum}, \bibinfo{person}{Michael Isard}, \bibinfo{person}{Josef Sivic}, {and}
  \bibinfo{person}{Andrew Zisserman}.} \bibinfo{year}{2007}\natexlab{}.
\newblock \showarticletitle{Object retrieval with large vocabularies and fast
  spatial matching}. In \bibinfo{booktitle}{\emph{{CVPR}}}.
\newblock


\bibitem[Qu et~al\mbox{.}(2020)]%
        {DBLP:conf/mm/Qu0CN020/campera}
\bibfield{author}{\bibinfo{person}{Leigang Qu}, \bibinfo{person}{Meng Liu},
  \bibinfo{person}{Da Cao}, \bibinfo{person}{Liqiang Nie}, {and}
  \bibinfo{person}{Qi Tian}.} \bibinfo{year}{2020}\natexlab{}.
\newblock \showarticletitle{Context-Aware Multi-View Summarization Network for
  Image-Text Matching}. In \bibinfo{booktitle}{\emph{ACM Multimedia}}.
\newblock


\bibitem[Rao et~al\mbox{.}(2021)]%
        {DBLP:conf/cikm/RaoQQW0021}
\bibfield{author}{\bibinfo{person}{Jun Rao}, \bibinfo{person}{Tao Qian},
  \bibinfo{person}{Shuhan Qi}, \bibinfo{person}{Yulin Wu},
  \bibinfo{person}{Qing Liao}, {and} \bibinfo{person}{Xuan Wang}.}
  \bibinfo{year}{2021}\natexlab{}.
\newblock \showarticletitle{Student Can Also be a Good Teacher: Extracting
  Knowledge from Vision-and-Language Model for Cross-Modal Retrieval}. In
  \bibinfo{booktitle}{\emph{{CIKM}}}.
\newblock


\bibitem[Ren et~al\mbox{.}(2017)]%
        {FASTERRCNN}
\bibfield{author}{\bibinfo{person}{Shaoqing Ren}, \bibinfo{person}{Kaiming He},
  \bibinfo{person}{Ross~B. Girshick}, {and} \bibinfo{person}{Jian Sun}.}
  \bibinfo{year}{2017}\natexlab{}.
\newblock \showarticletitle{Faster {R-CNN:} Towards Real-Time Object Detection
  with Region Proposal Networks}.
\newblock \bibinfo{journal}{\emph{TPAMI}} \bibinfo{volume}{39},
  \bibinfo{number}{6} (\bibinfo{year}{2017}), \bibinfo{pages}{1137--1149}.
\newblock


\bibitem[Robertson and Zaragoza(2009)]%
        {bm25}
\bibfield{author}{\bibinfo{person}{Stephen~E. Robertson} {and}
  \bibinfo{person}{Hugo Zaragoza}.} \bibinfo{year}{2009}\natexlab{}.
\newblock \showarticletitle{The Probabilistic Relevance Framework: {BM25} and
  Beyond}.
\newblock \bibinfo{journal}{\emph{Found. Trends Inf. Retr.}}
  \bibinfo{volume}{3}, \bibinfo{number}{4} (\bibinfo{year}{2009}),
  \bibinfo{pages}{333--389}.
\newblock


\bibitem[Robinson et~al\mbox{.}(2021)]%
        {DBLP:conf/iclr/RobinsonCSJ21/hnm}
\bibfield{author}{\bibinfo{person}{Joshua~David Robinson},
  \bibinfo{person}{Ching{-}Yao Chuang}, \bibinfo{person}{Suvrit Sra}, {and}
  \bibinfo{person}{Stefanie Jegelka}.} \bibinfo{year}{2021}\natexlab{}.
\newblock \showarticletitle{Contrastive Learning with Hard Negative Samples}.
  In \bibinfo{booktitle}{\emph{{ICLR}}}.
\newblock


\bibitem[Schuster and Paliwal(1997)]%
        {DBLP:journals/tsp/SchusterP97/bigru}
\bibfield{author}{\bibinfo{person}{Mike Schuster} {and}
  \bibinfo{person}{Kuldip~K. Paliwal}.} \bibinfo{year}{1997}\natexlab{}.
\newblock \showarticletitle{Bidirectional recurrent neural networks}.
\newblock \bibinfo{journal}{\emph{{IEEE} Trans. Signal Process.}}
  \bibinfo{volume}{45}, \bibinfo{number}{11} (\bibinfo{year}{1997}),
  \bibinfo{pages}{2673--2681}.
\newblock


\bibitem[Sharma et~al\mbox{.}(2018)]%
        {DBLP:conf/acl/SoricutDSG18/gcc}
\bibfield{author}{\bibinfo{person}{Piyush Sharma}, \bibinfo{person}{Nan Ding},
  \bibinfo{person}{Sebastian Goodman}, {and} \bibinfo{person}{Radu Soricut}.}
  \bibinfo{year}{2018}\natexlab{}.
\newblock \showarticletitle{Conceptual Captions: {A} Cleaned, Hypernymed, Image
  Alt-text Dataset For Automatic Image Captioning}. In
  \bibinfo{booktitle}{\emph{ACL}}.
\newblock


\bibitem[Su et~al\mbox{.}(2020)]%
        {vl-bert}
\bibfield{author}{\bibinfo{person}{Weijie Su}, \bibinfo{person}{Xizhou Zhu},
  \bibinfo{person}{Yue Cao}, \bibinfo{person}{Bin Li}, \bibinfo{person}{Lewei
  Lu}, \bibinfo{person}{Furu Wei}, {and} \bibinfo{person}{Jifeng Dai}.}
  \bibinfo{year}{2020}\natexlab{}.
\newblock \showarticletitle{{VL-BERT:} Pre-training of Generic
  Visual-Linguistic Representations}. In \bibinfo{booktitle}{\emph{ICLR}}.
\newblock


\bibitem[van~den Oord et~al\mbox{.}(2018)]%
        {DBLP:journals/corr/abs-1807-03748/Contrastive}
\bibfield{author}{\bibinfo{person}{A{\"{a}}ron van~den Oord},
  \bibinfo{person}{Yazhe Li}, {and} \bibinfo{person}{Oriol Vinyals}.}
  \bibinfo{year}{2018}\natexlab{}.
\newblock \showarticletitle{Representation Learning with Contrastive Predictive
  Coding}.
\newblock \bibinfo{journal}{\emph{CoRR}}  \bibinfo{volume}{abs/1807.03748}
  (\bibinfo{year}{2018}).
\newblock


\bibitem[Vaswani et~al\mbox{.}(2017)]%
        {DBLP:conf/nips/VaswaniSPUJGKP17}
\bibfield{author}{\bibinfo{person}{Ashish Vaswani}, \bibinfo{person}{Noam
  Shazeer}, \bibinfo{person}{Niki Parmar}, \bibinfo{person}{Jakob Uszkoreit},
  \bibinfo{person}{Llion Jones}, \bibinfo{person}{Aidan~N. Gomez},
  \bibinfo{person}{Lukasz Kaiser}, {and} \bibinfo{person}{Illia Polosukhin}.}
  \bibinfo{year}{2017}\natexlab{}.
\newblock \showarticletitle{Attention is All you Need}. In
  \bibinfo{booktitle}{\emph{NeurIPS}}.
\newblock


\bibitem[Velickovic et~al\mbox{.}(2018)]%
        {DBLP:conf/iclr/VelickovicCCRLB18/GAT}
\bibfield{author}{\bibinfo{person}{Petar Velickovic}, \bibinfo{person}{Guillem
  Cucurull}, \bibinfo{person}{Arantxa Casanova}, \bibinfo{person}{Adriana
  Romero}, \bibinfo{person}{Pietro Li{\`{o}}}, {and} \bibinfo{person}{Yoshua
  Bengio}.} \bibinfo{year}{2018}\natexlab{}.
\newblock \showarticletitle{Graph Attention Networks}. In
  \bibinfo{booktitle}{\emph{ICLR}}.
\newblock


\bibitem[Wang et~al\mbox{.}(2017)]%
        {DBLP:conf/iccv/WangZWLL17/Angular_loss}
\bibfield{author}{\bibinfo{person}{Jian Wang}, \bibinfo{person}{Feng Zhou},
  \bibinfo{person}{Shilei Wen}, \bibinfo{person}{Xiao Liu}, {and}
  \bibinfo{person}{Yuanqing Lin}.} \bibinfo{year}{2017}\natexlab{}.
\newblock \showarticletitle{Deep Metric Learning with Angular Loss}. In
  \bibinfo{booktitle}{\emph{{ICCV}}}.
\newblock


\bibitem[Wang et~al\mbox{.}(2019)]%
        {Wang_2019_ICCV/camp}
\bibfield{author}{\bibinfo{person}{Zihao Wang}, \bibinfo{person}{Xihui Liu},
  \bibinfo{person}{Hongsheng Li}, \bibinfo{person}{Lu Sheng},
  \bibinfo{person}{Junjie Yan}, \bibinfo{person}{Xiaogang Wang}, {and}
  \bibinfo{person}{Jing Shao}.} \bibinfo{year}{2019}\natexlab{}.
\newblock \showarticletitle{CAMP: Cross-Modal Adaptive Message Passing for
  Text-Image Retrieval}. In \bibinfo{booktitle}{\emph{ICCV}}.
\newblock


\bibitem[Wu et~al\mbox{.}(2021)]%
        {wu2021slua}
\bibfield{author}{\bibinfo{person}{Di Wu}, \bibinfo{person}{Liang Ding},
  \bibinfo{person}{Shuo Yang}, {and} \bibinfo{person}{Dacheng Tao}.}
  \bibinfo{year}{2021}\natexlab{}.
\newblock \showarticletitle{Slua: A super lightweight unsupervised word
  alignment model via cross-lingual contrastive learning}.
\newblock \bibinfo{journal}{\emph{arXiv preprint}} (\bibinfo{year}{2021}).
\newblock


\bibitem[Wu et~al\mbox{.}(2019)]%
        {DBLP:conf/mm/WuWSH19/saem}
\bibfield{author}{\bibinfo{person}{Yiling Wu}, \bibinfo{person}{Shuhui Wang},
  \bibinfo{person}{Guoli Song}, {and} \bibinfo{person}{Qingming Huang}.}
  \bibinfo{year}{2019}\natexlab{}.
\newblock \showarticletitle{Learning Fragment Self-Attention Embeddings for
  Image-Text Matching}. In \bibinfo{booktitle}{\emph{ACM Multimedia}}.
\newblock


\bibitem[Xu et~al\mbox{.}(2021)]%
        {vitae}
\bibfield{author}{\bibinfo{person}{Yufei Xu}, \bibinfo{person}{Qiming Zhang},
  \bibinfo{person}{Jing Zhang}, {and} \bibinfo{person}{Dacheng Tao}.}
  \bibinfo{year}{2021}\natexlab{}.
\newblock \showarticletitle{Vitae: Vision transformer advanced by exploring
  intrinsic inductive bias}.
\newblock \bibinfo{journal}{\emph{NeurIPS}} (\bibinfo{year}{2021}).
\newblock


\bibitem[Young et~al\mbox{.}(2014)]%
        {DBLP:journals/tacl/YoungLHH14}
\bibfield{author}{\bibinfo{person}{Peter Young}, \bibinfo{person}{Alice Lai},
  \bibinfo{person}{Micah Hodosh}, {and} \bibinfo{person}{Julia Hockenmaier}.}
  \bibinfo{year}{2014}\natexlab{}.
\newblock \showarticletitle{From image descriptions to visual denotations: New
  similarity metrics for semantic inference over event descriptions}.
\newblock \bibinfo{journal}{\emph{TACL}}  \bibinfo{volume}{2}
  (\bibinfo{year}{2014}), \bibinfo{pages}{67--78}.
\newblock


\bibitem[Yu et~al\mbox{.}(2021)]%
        {yu2021deep}
\bibfield{author}{\bibinfo{person}{Jun Yu}, \bibinfo{person}{Hao Zhou},
  \bibinfo{person}{Yibing Zhan}, {and} \bibinfo{person}{Dacheng Tao}.}
  \bibinfo{year}{2021}\natexlab{}.
\newblock \showarticletitle{Deep Graph-neighbor Coherence Preserving Network
  for Unsupervised Cross-modal Hashing}. In \bibinfo{booktitle}{\emph{AAAI}}.
\newblock


\bibitem[Zan et~al\mbox{.}(2022)]%
        {zan2022bridging}
\bibfield{author}{\bibinfo{person}{Changtong Zan}, \bibinfo{person}{Liang
  Ding}, \bibinfo{person}{Li Shen}, \bibinfo{person}{Yu Cao},
  \bibinfo{person}{Weifeng Liu}, {and} \bibinfo{person}{Dacheng Tao}.}
  \bibinfo{year}{2022}\natexlab{}.
\newblock \showarticletitle{Bridging Cross-Lingual Gaps During Leveraging the
  Multilingual Sequence-to-Sequence Pretraining for Text Generation}. In
  \bibinfo{booktitle}{\emph{arXiv preprint}}.
\newblock


\bibitem[Zhan et~al\mbox{.}(2018)]%
        {zhan2018comprehensive}
\bibfield{author}{\bibinfo{person}{Yibing Zhan}, \bibinfo{person}{Jun Yu},
  \bibinfo{person}{Zhou Yu}, \bibinfo{person}{Rong Zhang},
  \bibinfo{person}{Dacheng Tao}, {and} \bibinfo{person}{Qi Tian}.}
  \bibinfo{year}{2018}\natexlab{}.
\newblock \showarticletitle{Comprehensive distance-preserving autoencoders for
  cross-modal retrieval}. In \bibinfo{booktitle}{\emph{ACM Multimedia}}.
\newblock


\bibitem[Zhang et~al\mbox{.}(2020a)]%
        {DBLP:conf/emnlp/ZhangHJIS20}
\bibfield{author}{\bibinfo{person}{Bowen Zhang}, \bibinfo{person}{Hexiang Hu},
  \bibinfo{person}{Vihan Jain}, \bibinfo{person}{Eugene Ie}, {and}
  \bibinfo{person}{Fei Sha}.} \bibinfo{year}{2020}\natexlab{a}.
\newblock \showarticletitle{Learning to Represent Image and Text with
  Denotation Graph}. In \bibinfo{booktitle}{\emph{EMNLP}}.
\newblock


\bibitem[Zhang et~al\mbox{.}(2021)]%
        {vinvl}
\bibfield{author}{\bibinfo{person}{Pengchuan Zhang}, \bibinfo{person}{Xiujun
  Li}, \bibinfo{person}{Xiaowei Hu}, \bibinfo{person}{Jianwei Yang},
  \bibinfo{person}{Lei Zhang}, \bibinfo{person}{Lijuan Wang},
  \bibinfo{person}{Yejin Choi}, {and} \bibinfo{person}{Jianfeng Gao}.}
  \bibinfo{year}{2021}\natexlab{}.
\newblock \showarticletitle{VinVL: Revisiting Visual Representations in
  Vision-Language Models}. In \bibinfo{booktitle}{\emph{CVPR}}.
\newblock


\bibitem[Zhang et~al\mbox{.}(2020b)]%
        {DBLP:conf/cvpr/caan}
\bibfield{author}{\bibinfo{person}{Qi Zhang}, \bibinfo{person}{Zhen Lei},
  \bibinfo{person}{Zhaoxiang Zhang}, {and} \bibinfo{person}{Stan~Z. Li}.}
  \bibinfo{year}{2020}\natexlab{b}.
\newblock \showarticletitle{Context-Aware Attention Network for Image-Text
  Retrieval}. In \bibinfo{booktitle}{\emph{CVPR}}.
\newblock


\bibitem[Zhang et~al\mbox{.}(2022)]%
        {vitaev2}
\bibfield{author}{\bibinfo{person}{Qiming Zhang}, \bibinfo{person}{Yufei Xu},
  \bibinfo{person}{Jing Zhang}, {and} \bibinfo{person}{Dacheng Tao}.}
  \bibinfo{year}{2022}\natexlab{}.
\newblock \showarticletitle{ViTAEv2: Vision Transformer Advanced by Exploring
  Inductive Bias for Image Recognition and Beyond}.
\newblock \bibinfo{journal}{\emph{arXiv preprint}} (\bibinfo{year}{2022}).
\newblock


\end{thebibliography}

\end{document}